\documentclass[entropy,article,accept,moreauthors,pdftex]{mdpi} 

\usepackage{braket}
\usepackage{color}
\newcommand{\ZT}[1]{\textquotedblleft#1\textquotedblright}%
\newcommand{\dif}{\mathrm{d}}%
\newcommand{\pdif}[2]{\frac{\partial#1}{\partial#2}}%
\newcommand{\Nabla}{\vec{\nabla}}%
\newcommand{\Tr}{\operatorname{Tr}}%

\usepackage{seqsplit}
\newcommand{\WWW}[1]{\href{#1}{\seqsplit{#1}}}

\firstpage{1} 
\pubvolume{25}
\issuenum{2}
\articlenumber{315}
\pubyear{2023}
\copyrightyear{2023}
\externaleditor{Academic Editors: Antonio M. Scarfone, Christopher Jeynes, Michael Parker and Luisberis Velazquez
}
\datereceived{13 December 2022} 
\daterevised{15 January 2023} 
\dateaccepted{3 February 2023} 
\datepublished{8 February 2023} 
\hreflink{https://doi.org/10.3390/e25020315} 
\pdfoutput=1 
\pdfoutput=1

\Title{Thermodynamics of an Empty Box}

\TitleCitation{Thermodynamics of an Empty Box}

\Author{Georg J. Schmitz 
 $^{1,*}$\orcidA{}, Michael te Vrugt 
 $^{2}$\orcidB{}, Tore Haug-Warberg $^3$, Lodin Ellingsen $^3$, Paul Needham $^4$ \linebreak  and Raphael Wittkowski $^2$\orcidC{}}

\AuthorNames{Georg J. Schmitz, Michael te Vrugt, Tore Haug-Warberg, Lodin Ellingsen, Paul Needham, and Raphael Wittkowski}

\AuthorCitation{{Schmitz, G.J.;} 
 te Vrugt, M.; Haug-Warberg, T.; Ellingsen, L.; Needham, P.; Wittkowski, R.}

\address{%
	$^{1}$ \quad MICRESS Group, ACCESS e.V., Intzestr. 5, D-52072 Aachen, Germany\\
	$^{2}$ \quad Institut f\"ur Theoretische Physik, Center for Soft Nanoscience, Westf\"alische Wilhelms-Universit\"at M\"unster, D-48149 M\"unster, Germany; {michael.tevrugt@uni-muenster.de (M.t.V.); \linebreak  raphael.wittkowski@uni-muenster.de (R.W.)}
	\\
	$^{3}$ \quad Department of Chemical Engineering, Norwegian University of Science and Technology (NTNU),\linebreak   N-7491 Trondheim, Norway; {tore.haug-warberg@ntnu.no (T.H.-W.); lodin.ellingsen@ntnu.no (L.E.)}\\
	$^{4}$ \quad Department of Philosophy, University of Stockholm, SE-106 91 Stockholm, Sweden; {paul.needham@philosophy.su.se}\\}

\corres{Correspondence: g.j.schmitz@access-technology.de}

\abstract{A gas in a box is perhaps the most important model system studied in thermodynamics and statistical mechanics. Usually, studies focus on the gas, whereas the box merely serves as an idealized confinement. The present article focuses on the box as the central object and develops a thermodynamic theory by treating the geometric degrees of freedom of the box as the degrees of freedom of a thermodynamic system. Applying standard mathematical methods to the thermodynamics of an empty box allows equations with the same structure as those of cosmology and classical and quantum mechanics to be derived. The simple model system of an empty box is shown to have interesting connections to classical mechanics, special relativity, and quantum field theory.}


\begin{document}

	\section{Introduction}
The concepts of thermodynamics, general relativity, and quantum mechanics are considered the three main pillars of contemporary physics by the general physics community. Many philosophers and physicists, on the other hand, consider thermodynamics a phenomenological theory, which might be reducable to microscopic theories such as statistical mechanics. The question of whether or not thermodynamics is a fundamental theory is discussed from a philosophical perspective in a separate article~\cite{teVrugt2022}. The three pillars above jointly describe most of the essential features of our world such as ``\textit{{discrete}} things or states'' and their \textit{{interaction}} (quantum mechanics), the \textit{{direction}} of processes and their final target state---the \textit{{equilibrium}} (thermodynamics), and the relative \textit{{motion and interaction}} and, to some extent, \textit{{shape or change of shape}} of objects such as length contraction (special and general relativity). Quantum mechanics and general relativity both require the notion of space and time, whereas classical thermodynamics has so far only addressed space as a hidden variable, for example, as the ``volume'' in the ideal gas equation, and has only implicit time dependence, e.g., in the ``initial'' and ``final'' states. Basically, thermodynamics describes the \textit{{most probable state}} of a complex system allowing for multiple states that can be realized under given conditions. This can be seen when connecting it to microscopic physics via statistical mechanics, where the equilibrium state (which is what thermodynamics describes) corresponds to the most probable state of a system under certain conditions. Thermodynamics is, however, broader than just the statistical description linked to it. In this article, we consider equilibrium thermodynamics.
 
 \textit{{Probability theory}} has  been claimed as being the logic of science \cite{Jaynes}, which highlights the relation between the notions of entropy in thermodynamics and information theory. Two further important pillars of physics are geometry, which addresses the \textit{{shape}} of objects, and algebra, which allows for the quantification of both the number of objects and their individual properties. Geometry, algebra, and, eventually, their combination in terms of geometric algebra \cite{GeometricAlgebra} (not addressed here) form the basis of many other theories.
	
	\textit{{Thermodynamics}} plays an important role in physics and often seems to be overlooked or at least not adequately interpreted or considered. A number of notions used in quantum physics and cosmology draw on thermodynamic notions. Some examples are the ``entropy'' of a black hole \cite{Bekenstein1972, Schmitz2018}, the photon ``gas'' \cite{planck1915}, the ``temperature'' of the universe \cite{Turner1993}, the ``adiabatic expansion'' of the universe \cite{GorbunovRubakov2011, Lemaitre1927, Friedman1922}, and ``entropic'' gravity \cite{Verlinde2011}.
	
	\textit{{Geometry}} also plays a role in many phenomena in physics. Some examples are Kepler's  laws on the ``ellipsoidal'' motion of the planets \cite{Kepler}, the Schwarzschild ``radius'' \cite{Schwarzschild}, the ``curvature'' of spacetime \cite{Einstein}, the ``area'' of holographic projection \cite{t'Hooft}, confinement in a ``volume'' as a classical quantization example \cite{ParticleInABox} including black-body radiation \cite{Planck1930}, the ``dimensional confinement'' as the origin of Giant Magneto Resistance \cite{Gruenberg}, or the ``droplet model'' of the atomic nucleus \cite{Weizsaecker}.
	
	An important \textit{{relation between thermodynamics and geometry}} of a physical object is provided by the entropy equation of a black hole. The dimensionless (i.e., without Boltzmann constant $k_B$) entropy $S$ relates to the surface area of the black hole via the famous Bekenstein--Hawking equation \cite{Bekenstein19xx}:	\begin{equation}
		S=\frac {A}{4l_p^2}   \enspace \text{(geometric formulation)}, 
		\label{geometric}
	\end{equation}
	\begin{equation}
		S=\frac{c^3A}{4\hbar G}  \enspace \text{(physics formulation)}.
		\label{physics}
	\end{equation}
		Here, $A$ is the surface area of the event horizon, $c$ is the speed of light, $\hbar$ is the reduced Planck constant, and $G$ is the gravitational constant. Obviously, Equation \eqref{geometric} follows from\linebreak   Equation~\eqref{physics} if we define the Planck length as $l_p=\sqrt{G\hbar/c^3}$. Recently, it has been shown that the geometric formulation of Equation \eqref{geometric} can be derived on the basis of mere geometric/statistical considerations for a sphere having a boundary with a finite thickness \cite{Schmitz2018}. The present article aims to broaden this ``geometric thermodynamics'' approach toward a fully anisotropic geometry of an empty box.
	
	\subsection{Scope}
	The aim of the present article is to connect the elementary thermodynamics of an empty box with fundamental physical theories, giving us, to coin a phrase, “physics out of a box”. Thermodynamics, considered from the outset as a theory in its own right, will be combined with geometry. This link has not yet been properly established, mainly because thermodynamics does not embody a length scale. This deficiency can be addressed by assuming that interfaces have a finite thickness, which then represents a length scale in a thermodynamic system \cite{Schmitz2003}. Another obvious link between thermodynamics and geometry is the concept of volume. Thermodynamics treats the volume of a system simply as a parameter, whereas geometry allows its calculation from more basic properties. The description of an anisotropic volume---a box---will be investigated in detail in the present article, treating any parameter occurring in this description as a potential degree of freedom of the thermodynamic system. The present article focuses on exploiting the scalar thermodynamic potentials at the mathematical foundations of thermodynamics. 
	
	The primary concern is to investigate what can be learned from this approach. Expectations range from a possible description of 
 relations between forces and temperature as found, e.g., in the Unruh temperature \cite{Unruh}, to special relativity and perhaps further unexpected terms and equations. As will be shown, applying elementary operations to a thermodynamic model for the box results in several equations that formally resemble known equations used in special and general relativity, as well as quantum mechanics. Analogies between general relativity and condensed matter systems \cite{BarceloLV2011} are known to be useful for investigating gravitational systems, and finding analogies between systems from classical statistical mechanics and quantum mechanics is a rapidly growing field of research \cite{teVrugtFHHTW2022}. Moreover, the successes of entropic gravity \cite{Verlinde2011} indicate that the mathematical relation between thermodynamics and general relativity might not be merely an analogy, but reflect a deeper connection.

	\subsection{Outline}

	Starting with the above introductory motivation highlighting the role of thermodynamic and geometric concepts in physics, a short historical review of thermodynamics is followed by an introduction to \textit{{thermodynamic potentials}} and their mutual convertibility via Legendre transformations. The internal energy $U(S,V,N)$ is selected as the thermodynamic potential for further discussion in the present article. The three \textit{{state variables $S$ (entropy), $V$ (volume), and $N$ (particle number, not treated here)}} are then briefly discussed before taking a closer look at the role of the volume state variable $V$. In classical thermodynamics, this state variable goes along with a change in the \textit{{volume value}}. 
	The present article will discuss \textit{{changes in volume shape}} while the volume value is kept constant. The discussion is based on a box, and an entire ensemble of boxes with the same volume values can easily be imagined. One of the major differences between such boxes, however, is their anisotropy.
			To allow for comprehensive treatment and a comparison of ``areas'' and ``volumes'' in a common description, a concept of \textit{{quantized space}} is introduced. This concept treats volumes as consisting of an \textit{interior volume} and a \textit{{boundary volume}}. \textit{{Boundary volume}} and  \textit{{interior volume}} thus represent two degrees of freedom in the thermodynamic system, which can be varied independently under the constraint of their sum being a constant volume value. {For this reason, our article might also be called \textit{Thermodynamics of constraints}. We have nevertheless chosen to call it \textit{Thermodynamics of an empty box} to emphasize that we focus on the special case of a box.} A variety of dimensionless entities can be defined based on {our} quantization scheme and examples of their exploitation will be provided for the gravitational potential and black-hole entropy. Uniaxially \textit {{squeezing the box}} leads to insights concerning special relativity. \textit{{Translating the box}} will be shown to result in Newtonian mechanics and an expression for the Unruh temperature. The summary in the conclusion provides an outlook on (i) \textit{{filling the box}} with particles possibly leading to \textit{{black-body radiation}} and classical quantum mechanics of a \textit{{particle in a box}}, \linebreak  (ii) \textit{{oriented box surfaces}} as a possible link to electrostatics, and (iii) \textit{{higher-order correlations}} as a possible pathway to other fundamental forces.

\section{Thermodynamics}
	The connotation of the term “thermodynamics” has changed somewhat since the creation of the theory around 1850 and is {not really concerned with the dynamics of heat.} Etymologically, the term “thermodynamics” derives from the Greek words “thermé” (heat) and “dynamics” (force). The concept of force was for a long time not distinguished from what came to be called energy, and the term “thermodynamics” in contemporary usage thus has the sense of “thermal energy”. The perception of the science of thermodynamics in different communities is heterogeneous with a variety of perspectives.
	
	There is a \textit{{classical perspective}} on thermodynamics that is typically taught at university. It relates, e.g., to heat and work or Carnot cycles and originates from the age of the steam engines.
	
	The \textit {{chemical thermodynamics perspective}} relates to the use of thermodynamics to predict energy and matter exchanges that occur in chemical reactions during phase changes or during the formation of solutions. The work of Josiah Willard Gibbs \cite{JWGibbs,JWGibbs2,GibbsElementary,Guggenheim} on the applications of thermodynamics was instrumental in transforming physical chemistry into a rigorous inductive science. The following state functions and thermodynamic potentials are of primary concern in chemical thermodynamics: the internal energy ($U$), the enthalpy ($H$), the entropy ($S$), and the Gibbs free energy ($G$). These thermodynamic potentials also form the basis for the fundamental \textit{{mathematical perspective}}.
	
	The \textit{{statistical mechanics perspective}} essentially goes back to Ludwig Boltzmann, who developed the fundamental interpretation of entropy in terms of a collection of microstates; James Clerk Maxwell, who developed models of probability distribution of such states such as the Maxwell distribution of velocities; and Josiah Willard Gibbs, who coined the name ``statistical mechanics'' for the field in 1884 \cite{Gibbs}.
	
	The \textit{{computational thermodynamics}} perspective has evolved from the use of thermodynamic data for the optimization of materials and processes. Computational thermodynamics is generally based on the CALPHAD method \cite{SundmannLukasFries} and makes use of thermodynamic databases. These are not simple databases and have underlying models such as a polynomial description of the Gibbs energies of the various phases in an alloy system. Nowadays, numerous thermodynamic databases and related software tools (e.g., \cite{ThermoCalc,FactSage,Pandat,JMatPro}) are regularly used to develop and optimize innovative materials and their processing.
	
	The \textit{{resource/control theory perspective}} (see \cite{Wallace2014,Myrvold2020b,Myrvold2021}) considers thermodynamics to be a theory that describes possible manipulations of a system and the response of the system to these manipulations. These introduce a heat/work distinction. 
	
	The \textit{{fundamental mathematical}} perspective encompasses the mathematical foundations of thermodynamics such as the Euler homogeneity principle, thermodynamic potentials, and Legendre transformations \cite{Callen1991}, allowing switching between the different potentials.
	
	The \textit{{geometric perspective}} has been developed during recent decades. In particular, the phase-field method (for a review see, e.g., \cite{PhaseField}) and its implementation in contemporary software tools (e.g., \cite{MICRESS}) allows for a spatially resolved description of shapes and structures and their evolution, even in complex technical alloys \cite{SchmitzTechAlloys}. Further investigations in the \textit{geometric perspective} relate to the use of entropic concepts for the description of shapes (e.g., \cite{ShapeEntropy}) or the description of boundaries and higher-order junctions by mereotopology \cite{Schmitz2022}. It also seems interesting to note the similarity between Gibbs' phase rule $F=C+P-2$  (degrees of \underline{f}reedom $F$, number of \underline{c}onstituents $C$, number of \underline{p}hases $P$) and the Euler characteristic for convex polyhedra $V-E+F=2$ (number of \underline{v}ortices $V$, number of \underline{e}dges $E$, number of \underline{f}aces $F$). This similarity is not coincidental but is already apparent in Gibbs' ``Graphical Methods in the Thermodynamics of Fluids'' and a ``Method of Geometrical Representation of the Thermodynamic Properties of Substances by means of Surfaces'' \cite{GibbsGeometrical,GibbsGraphical}.
	
	The \textit{{physics perspective}} on thermodynamics goes far beyond the classical thermodynamics of steam engines. Similar to the chemical thermodynamics perspective, which describes chemical phenomena in terms of thermodynamics, the physics perspective aims at describing phenomena of physics based on thermodynamic principles. This perspective encompasses notions such as black-hole entropy \cite{Bekenstein1972}, entropic gravity \cite{Verlinde2011}, and others. 
	
	The present article is especially concerned with the last three perspectives.
	
	\subsection{Mathematical Thermodynamics}
	
	From a mathematical perspective, a thermodynamic system is any system---real or not---that fulfills Euler's homogeneity principle in addition to the axioms of equilibrium~\cite{Callen1991}.\label{eulerh} This theorem assumes that the following condition defines the function $F(X)$ as homogeneous of the order $k$:
	\begin {equation}
	F(X) = \lambda^k f(x); \quad X := \lambda x; \quad \lambda \in \mathbb{R}.
 \label{FXfx}
	\end {equation}
    The free variables are assumed to be $x$ without saying anything about their true nature, except that they satisfy the scaling law $\lambda x$ including $\lambda = 0$, which corresponds to $F=0$ regardless of the physical reference point.
    The functions $F(X)$ and $f(x)$ represent two different scales of the thermodynamic state but otherwise share the same function definition (see Equation \eqref{FXfx}).
    This is in contrast to, say, the functions $U(S,V,N)$ and $U(T,V,N)$, which are (typically) used to express the same state but with two different function expressions.
	All thermodynamic potentials for extensive systems must be homogeneous functions of order $k=1$. For a function with several variables $x_i$, the integration theorem of Euler~\cite{Ram2009}~yields
	
	\begin {equation}
	X_i\pdif{F}{X_i} = k F(\lambda\mathbf{x}),
	\label{eulermultiple}
	\end {equation}
	where $\mathbf{x}$ stands for all variables $x_i$ and the Einstein summation convention is applied.
	There are many interesting side effects of homogeneity of which only three are mentioned here.
	First, we note that because $F \equiv f$, Equation \eqref{eulermultiple} also applies to $f (\mathbf{x})$.
	Second, we note that a homogeneous function is identified by its partial derivatives. The exception is the intensive function with $k=0$, which is important because, although $F$ is determined by $\pdif{F}{X_i}$, the first derivatives are not determined by their first derivatives. This is because the homogeneity of $F$ (and its derivatives) is reduced upon differentiation. In fact, if $F$ is homogeneous of order $k=1$, $\pdif{F}{X_i}$ is homogeneous of order $k=0$, and so on. 
	Third and finally, a direct consequence of Equation \eqref{eulermultiple} and the differentiation rule just mentioned is 
	\begin {equation}
	X_j \pdif{^2 F}{X_j\partial X_i}=0 \quad \forall i
	\end {equation}
	which, in thermodynamics, is more commonly written in the differential form
	\begin {equation}
	X_i\,\dif\left(\pdif{F}{X_i}\right) = 0
	\label{FormalGibbsDuhemEquation}
	\end {equation}
	and is then called the Gibbs--Duhem equation.
	
	\subsubsection{\label{TDPotentials}Thermodynamic
	Potentials}
	Thermodynamic potentials are abstract functionals in several variables. These variables are not always easy to determine because a thermodynamic system can have more internal degrees of freedom than external ones (sometimes called work modes). A simple example would be the chemical reaction of two chemical species $\mathrm{A}$ and $\mathrm{B}$, usually written as $\mathrm{A} \Leftrightarrow \mathrm{B}$. Here, only the total amount $\mathrm{A} + \mathrm{B}$ can be varied externally, whereas the distribution of $\mathrm{A}$ in relation to $\mathrm{B}$ represents an internal degree of freedom that rests on the principle of thermodynamic equilibrium.
	The extremum of a thermodynamic potential allows the specification of a special state called equilibrium, corresponding to, e.g., a minimum of energy or maximum of entropy. A specific example of a thermodynamic potential is the internal energy $U$, which provides the basis for all discussions throughout the present article. It is a function of the three extensive variables entropy $S$, volume $V$, and number of particles $N$
	\begin{equation}
		U=U(S,V,N).
		\label{Upotential}
	\end{equation}
	The Euler homogeneity theorem stated in Equation \eqref{eulermultiple} applied to $U$ requires
	\begin{equation}
		U=\pdif{U}{V} V + \pdif{U}{S}S+ \pdif{U}{N} N.
		\label{U_Euler}
	\end{equation}
	Introducing the definitions for pressure $p$, temperature $T$, and chemical potential $\mu$
	\begin{equation}
		\pdif{U}{V}=:-p, \enspace \enspace \pdif{U}{S}=:T, \enspace \text{and} \enspace  \pdif{U}{N}=: \mu 
		\label{pTmuDefinitions}
	\end{equation}
	allows the internal energy potential $U$ to be formulated as
	\begin{equation}
		U = TS + (-p)V + \mu N.
		\label{InternalEnergy}
	\end{equation}
	\textls[-15]{By utilizing the definitions in Equation \eqref{pTmuDefinitions}, the total differential of $U$ in Equation \eqref{Upotential} reads:}
	\begin{equation}
		\dif U = T\dif S + (-p)\dif V + \mu\dif N. 
  \label{dU_equation}
   \end{equation}
Since $U$ is a Euler homogeneous function, a second non-trivial differential arises if the differential of Equation \eqref{InternalEnergy} is instead used as a starting point for the differentiation. This is the celebrated Gibbs--Duhem equation from Equation \eqref{FormalGibbsDuhemEquation}:
	\begin{equation}
		S\dif T + V\dif(-p) + N\dif\mu = 0.
		\label{GibbsDuhemEquation}
	\end{equation}
	As a representative of Euler homogeneity in differential form, this equation must be fulfilled everywhere and at all times for a thermodynamic continuum.
	{For it to be valid in the case of a nonequilibrium system whose state changes over time, one has to assume that the size of the system in the form of $S, V, N$ is small enough so that the respective densities are constant throughout the volume and that the changes in the system are slow enough so that the intensive state variables retain their macroscopic significance. Even in a spatially inhomogeneous system, thermodynamic concepts often continue to be applicable in the case of \ZT{local equilibrium}, where an equilibrium-like description holds within each small volume element.}

	Both Equations \eqref{dU_equation}  and \eqref{GibbsDuhemEquation} are invariant under translations/rotations in space and time. Such symmetries are significant because of their role in Noether's theorem \cite{Noether}, which has recently also been applied to thermodynamic potentials \cite{HermannS2021}. 
	Thermodynamic potential formulations (see Appendix \ref{Legendre}) comprising macroscopically and practically measurable state properties such as pressure and temperature and countable or measurable quantities such as particle number and volume can be derived in this way, with an example being the Gibbs free energy $U_{SV} =: G(T,-p,N)$. 
	In summary, all thermodynamic potentials can be transformed into each other and any one of them is, accordingly, sufficient for describing a thermodynamic system. The present article considers thermodynamics to be a theory in its own right \cite{teVrugt2022} and discusses the internal energy $U$ of an empty box, starting with a discussion of the classical internal energy $U=U(S,V,N)$.
\subsection{Scalar Potentials, Gradients, and Forces\label{spgf}}

Since the internal energy potential formulation depends neither on space nor time, it holds for all positions $\vec r$ and times $t$. We now assume that the thermodynamic potentials can depend on $\vec r$ and $t$. For an empty box ($N=0$), Equation \eqref{Upotential} then reads
\begin{equation}
		U (\vec r, t) = TS(\vec r, t) + (-p)V(\vec r, t)\text{ for all positions $\vec r$ and times $t$}.
		\label{InternalEnergyField}
\end{equation}
A possible physical reason why $S$ and $V$ depend on $\vec{r}$ and $t$---a dependence not usually assumed in thermodynamic treatments---is that the system is in a spatially or temporally changing environment, something that is, in fact, the rule rather than the exception for actual physical systems. For example, if a gas confined by a piston, which changes its volume depending on the external pressure, is in an environment where the external pressure changes with position or time, then the volume and entropy will also change. In this work, we are not concerned with a specific physical example of this kind. (Otherwise, it could affect our choice of thermodynamic potential---for a gas with prescribed external pressure, the enthalpy or Gibbs free energy could be a more natural choice. This does not, however, change the reason for studying thermodynamic potentials in the form of Equation \eqref{InternalEnergyField}.) 
Instead, we consider a simple mathematical object (a box), assume it to have a certain internal energy, and explore the consequences.

The typical interpretation for the differential of {$U$} relates to a ``change'' between an \textit{{initial state}} and a \textit{{final state}} and does not consider any spatial dependencies:
\begin{equation}
		U (t) = TS(t) + (-p)V(t)\enspace \text {for all times $t$}.  
		\label{InternalEnergyContrast}
\end{equation}
The differential then reads
\begin{equation}
		\dif U  = T\dif S + (-p)\dif V \rightarrow \dot U= T\dot S +(-p)\dot V,  
		\label{difInternalEnergyContrast}
\end{equation}
where the overdot indicates a derivative with respect to time. For a stationary spatial pattern with no dependence on time, we obtain
\begin{equation}
		U (\vec r) = TS(\vec r) + (-p)V(\vec r)\enspace \text{for all 
		positions $\vec r$.}  
		\label{InternalEnergyContrastSpace}
\end{equation}
The differential then reads
\begin{equation}
		\dif U  = T\dif S + (-p)\dif V \rightarrow \vec\nabla U= T\vec\nabla S +(-p)\vec\nabla V.
		\label{difInternalEnergyContrastGradient}
\end{equation}
Note that we use the symbol $\vec\nabla$ to denote a derivative with respect to all dependencies on $\vec r$ and the symbol $\pdif{}{\vec r}$ for derivatives with respect to explicit dependencies on $\vec r$. For example, if we have a potential $U(V(\vec{r}),\vec{r})$, we have 
\begin{align} 
\Nabla U(V(\vec{r}),\vec{r})&= \pdif{U}{V}\pdif{V}{\vec{r}} + \bigg(\pdif{U}{\vec{r}}\bigg)_V,\\
\pdif{U}{\vec{r}} =& \bigg(\pdif{U}{\vec{r}}\bigg)_V,
\end{align}
with the subscript $V$ indicating (as is usual) that the derivative is taken at constant $V$. Note that there is, in this example, no difference between $\Nabla V$ and $\pdif{V}{\vec{r}}$. In a Hamiltonian system, a force is given by the negative gradient of the internal energy $U$:
	\begin{equation}
		\vec{F} = -\vec\nabla U.
		\label{forcedefinition}
	\end{equation}
Equation \eqref{forcedefinition} might not seem obvious since the thermodynamic potential $U$ is not the same mathematical object as the potential energy from classical mechanics and since it is the potential energy whose gradient gives the force. However, if we allow a thermodynamic potential to explicitly depend on the centroid position $\vec r$, we are essentially treating $\vec r$ as a relevant macroscopic variable. This was done in a microscopic treatment based on the Mori--Zwanzig formalism \cite{Mori1965,Zwanzig1960,teVrugtW2019,teVrugtW2019b} in Ref. \cite{CamargodlTDZEDBC2018} for the simpler case of a sphere immersed in a fluid. There, it was demonstrated that the thermodynamic conjugate of the centroid position---in this work, given by $\vec\nabla U$---can indeed be interpreted as the mechanical force. Here, we consider an explicit dependence of $U$ on $\vec r$---corresponding to treating $\vec r$ as a relevant variable---in Section \ref{newton}. Note that the definition in Equation \eqref{forcedefinition} is more general than the mechanical definition because it also includes entropic forces.

Assuming that $S$ and $V$ are position-dependent, we obtain a potential 
\begin{equation}
	U = U (S(\vec r),V (\vec r))
\end{equation}
and can calculate the gradients
\begin{equation}
	-\vec\nabla U = -T\vec\nabla S + p\vec\nabla V .
	\label{PotentialForces}
\end{equation}
These can be identified with an entropic force and kinetic force:
\begin{equation}
	\vec F_{\mathrm{total}} = \vec F_{\mathrm{entropic}} + \vec F_{\mathrm{kinetic}} .
\end{equation}
Imposing the condition $\vec F_{\mathrm{total}} =\vec 0 $ in line with the assumption of an isolated system with $\vec\nabla U =\vec{0}$ yields
\begin{equation}
 \vec F_{\mathrm{entropic}} =  - \vec F_{\mathrm{kinetic}} \enspace \text {(actio = reactio)}.
\end{equation}
This corresponds to one of the necessary conditions of mechanical equilibrium---the sum of all forces must vanish ($\vec F_{\mathrm{total}} = \vec 0$). {It can also be thought of as a manifestation of Newton's third law.} The other necessary condition is a vanishing sum of torques, which is not treated here.
A kinetic force can thus be defined as

\begin{equation}
		\vec F_{\mathrm{kinetic}} := -\pdif{U}{V}\pdif{V}{\vec r} = p\vec\nabla V \enspace \text {(kinetic force)}.
\end{equation}
An entropic force can accordingly be defined as
\begin{equation}
		\vec F_{\mathrm{entropic}} := -\pdif{U}{S}\pdif{S}{\vec r} = -T\vec\nabla S \enspace \text{(entropic force)}.
\end{equation}

The relation $\pdif{U}{S}=T$ is used here. This reasoning recovers the standard definition $\vec F = - T\vec \nabla S$ \cite{TreumannB2019} of the entropic force. The interesting fact here is that ``force''---up to some constant---is \textit{the gradient of entropy}.

\subsection{\label{ofanempty}The Internal Energy of an Empty Box}

Starting with the classical formulation for the internal energy of a thermodynamic system $U=U(S,V,N)$, it is obvious that the classical state variable $N$ takes the value $0$. The classical internal energy formulation thus reduces to $U=U(S,V)$. In classical thermodynamics, the volume $V$ is treated as a scalar, positive, and real value, denoting the size of the region within which a thermodynamic system is confined. Here, we define the volume $V$ of a box as
	\begin{equation}
		V=\vec{a}\cdot(\vec{b}\times\vec{c}).
		\label{spadeproduct}
	\end{equation}
with the vectors $\vec a$, $\vec b$, $\vec c $ denoting the vectors spanning the volume. For a box with mutually pairwise perpendicular axes, the spade product reduces to the simple product. Being the product of three vectors, however, the volume defined in Equation \eqref{spadeproduct} is a pseudoscalar because inverting the sign of each of the three vectors in a parity operation will change the sign of the product of these three vectors, i.e., the sign of the value of the volume as well. This change in sign corresponds to a change in helicity, i.e., a transition from a left-handed coordinate system to a right-handed one. Consequently, our definition of $V$ differs from the standard one in that it is a pseudoscalar and not a scalar. The motivation for this is that it allows, in principle, the treatment of the orientation of the faces of the box as an additional degree of freedom (see Section \ref{beyondtheempty}). However, throughout the present article, we do not consider the orientations of vectors and simply describe the volume as a function of their absolute values. This volume, which we assume (as is standard) to be positive, is given~by
	\begin{equation}
		\vert V\vert = V = abc.
	\end{equation}
where $a$,$b$,$c$ are the absolute values of the vectors $\vec a$,$\vec b$,$\vec c $. A precise formulation of the internal energy is, therefore, $U=U(S,\vert V \vert)$. We maintain the interpretation and notation of $V:=\vert V \vert$. In most cases, we are interested in isochoric processes, i.e., processes that do not change the value of the volume of the box (i.e., $\dif V=0$). We can already note here that a variety of boxes exists, all having the same volume $V$ but differing in their anisotropy $\kappa$. The anisotropy can be defined, e.g., as the ratio of the a and b axes, keeping c constant. Squeezing a box thus generates the first new degree of freedom, $\kappa$, entering into the internal energy formulation:
	\begin{equation}
		U=U(S,\vert V \vert, \kappa, ...).
	\end{equation}	
We neglect any dependence on $N$ since here, we only consider an empty box. Finally, we write $V$ for $\vert V\vert$ to simplify the notation (and to keep it in the usual form) and include the centroid position $\vec r $. The internal energy of an empty box is, therefore, formulated as
	\begin{equation}
		U= U(S,V,\kappa,\vec r, ...) \enspace \text{(internal energy of a classical empty box)}
		\label{newinternalenergy}
	\end{equation}
with $V = abc$ being the scalar volume value, $\kappa$ being some anisotropy factor (see Section~\ref{squeezing} on \textit{{squeezing the box}}),  and $\vec r$ denoting the centroid position (see Section \ref{translating} on \textit{{translating the box}}{, where we also discuss the relation of this dependence to kinetic terms}). All these represent \textit{{independent}} DoFs (degrees of freedom) for a fixed absolute scalar value of the volume (i.e., $V = abc = constant$). They are independent since the box might, for example, be translated without being squeezed or be rotated without being translated, etc. The differential $\dif U$ then reads
	\begin{equation}
	\dif U = \pdif{U}{S}\dif S+ \pdif{U}{V}\dif V + \pdif{U}{\kappa}\dif\kappa + \pdif{U}{\vec r}\cdot \dif\vec r +......
	\end{equation}
{It is interesting to see how our considerations affect the Euler equation. We focus on position dependence. If $\vec r$ and $-\vec F$ are conjugate variables, \ZT{standard} thermodynamics would lead us to expect that this pair of variables also appears in the Euler equation, i.e., that it reads (ignoring $N$) $U= TS - pV - \vec{r}\cdot\vec{F}$. Since the Euler equation is derived from the assumption that $U$ is a first-order homogeneous function, whether this holds depends on whether $U(\lambda\vec{r}) = \lambda U(\vec{r})$ holds. This will be the case for a spatially homogeneous force field $\vec{F}$, where the potential is given by $U = -\vec{r}\cdot\vec{F}$. In general, however, this does not have to be the case since $\vec{F}$ may also depend on $\vec{r}$ (note that other homogeneity assumptions, such as $U(\lambda V) = \lambda U(V)$, may also break down).}

{ On the level of phenomenological equilibrium thermodynamics, a system (such as a gas) is, in general, simply \textit{{assumed}} to be described by a certain functional $U$ (for an isolated system) and to have a certain equation of state. Deriving such an equation of state from the microscopic properties of the system lies in the realm of statistical mechanics. What we have to assume about the box as a physical system is that it has a certain energy and that this energy depends on the parameters of the system. Constraining the box to have a certain volume or shape requires us to put energy into it, for example, by compressing, expanding, or deforming it. The pressure $p$ and temperature $T$ then arise as conjugate variables to volume $V$ and entropy $S$, respectively. If the box is not empty, for example, because it contains a gas, the internal energy $U$ of the system box + gas would also depend on the number of gas particles $N$. In this case, the temperature of the box would, in equilibrium, have to be the same as the temperature of the gas (making the reasonable assumption that heat can flow between the gas and box). Note that $N$ denotes only the number of gas particles, i.e., even if $N=0$ (no gas), the box, being a macroscopic physical object, still consists of a large number of particles and thus allows for a thermodynamic description.}

Equation \eqref{newinternalenergy} has been labeled the \textit{{internal energy of a classical empty box}} because it treats the \textit{{classical}} box as bound by a mathematically sharp interface. The thickness of the boundary or, in other words, the fraction $\Phi_B$ that the boundary volume takes of the total volume, however, represents another very interesting DoF for an empty box, which is discussed next, providing an intriguing view on entropy.

\subsubsection{\label{morethanstatistics}Entropy---More than Statistics}
	\textit{S} is the state variable for entropy. In the context of information theory, it is a dimensionless quantity \cite{Shannon}. Entropy can be defined in various ways. In physics, a familiar notion is the \textit{{Boltzmann entropy}} \cite{Frigg2008}, which measures the number of microstates corresponding to a given macrostate. The challenge  here is the identification of the possible ``microstates'' of an empty box. The macrostate of the box may be specified by its total volume. This volume may be composed of the interior volume of the box $V_I$ and its boundary volume $V_B$, which sum up to the total volume $V_{\mathrm{tot}}$. What this means for interpreting a boundary as a part of the volume is discussed in greater detail in Section \ref{chopping}. The fraction that the boundary volume takes of the total volume thus becomes a new DoF, which vanishes in the classical sharp interface thermodynamics, where the ``volume'' of the two-dimensional (2D) boundary would be exactly $0$. A macrostate given by a total volume can then be specified by summing the fraction of a system that belongs to the interior ($\Phi_I$) and the boundary ($\Phi_B$). Being fractions, they sum up to one:	\begin{equation}
		1 = {\Phi_I}+ {\Phi_B}.
		\label{sumphii}
	\end{equation}
	{Note that here, we are assuming that boundaries have a finite extension in the third dimension (see Section \ref{chopping}) such that both $\Phi_I$ and $\Phi_B$ are well-defined and have finite values. $\Phi_I$ and $\Phi_B$ will sum up to one if the system consists only of a boundary and interior, which is true by definition if we take the interior to be the part of the system that does not belong to the boundary.}
	Simply squaring Equation \eqref{sumphii} leads to 
	\begin{eqnarray}
		{\Phi_I}+ {\Phi_B}= {\Phi_I}^2+ {\Phi_B}^2 + 2\Phi_I\Phi_B
	\end{eqnarray}
	and so
	\begin{equation}
		\Phi_I(1-\Phi_I) +\Phi_B(1-\Phi_B)= 2\Phi_I\Phi_B.
		\label{entropyequation}
	\end{equation}
	Each of the two terms on the left-hand side of Equation \eqref{entropyequation} corresponds to the lowest order of the Taylor expansion of an entropy-type logarithmic formulation (for $\Phi_i\leq 1$) \cite{Bronstein}:
	\begin{equation}
		\Phi_i(1-\Phi_i)  \thickapprox - \Phi_i\ln\Phi_i.
	\end{equation}
	This mathematical similarity to expressions typically occurring in the definition of entropy motivates us {to}  define a dual-state entropy $S$ as 
		\begin{equation}
		S=2\Phi_I\Phi_B\approx -\sum\limits_{i=I,B}^{}{\Phi_i\ln\Phi_i} \enspace \enspace \text{(Dual-State Entropy Equation)}.
		\label{dualstate}
		\end{equation}
		Note that the logarithmic terms here approximate the exact expression $2\Phi_I\Phi_B$ and not vice versa.
		A similar argument is commonly used in the study of entanglement. If we want to know whether a quantum system described by a statistical operator $\rho$ is entangled, we can calculate its entropy $S=-\Tr(\rho\ln\rho)$ with the trace $\Tr$. This \ZT{entanglement entropy} vanishes for a pure state. Since this entropy is, in general, difficult to calculate, one can instead use the linear entropy $S_l=\Tr(\rho)-\Tr(\rho^2)$ \cite{HahnWK2020}. (The difference, of course, is that here, one would consider the linear entropy to be the approximate one. {Since we are not discussing quantum-mechanical entanglement here, the discussion of quantum entanglement is merely an analogy.})
	
	 {Although} we are free to define a quantity $S$ that we call \ZT{entropy} in any way we want, it is not generally guaranteed that this entropy is in any way related to a quantity that is usually thought of as \ZT{entropy} in thermodynamics (or statistical mechanics). Very roughly speaking, one would usually define the (Boltzmann) entropy as a measure for the number of microscopic realizations of a given macrostate and assume that systems maximize their entropy because this is the most likely behavior on statistical grounds. Thus, it would confirm our definition in Equation \eqref{dualstate} if it is found that the maximum of $S$ is, for a simple model system, indeed the one with the largest number of microscopic realizations. That this is indeed the case is demonstrated in Appendix \ref{boltzmann}. {In cases where the entropy in Equation \eqref{dualstate} coincides with the Boltzmann entropy, we can (assuming that the Boltzmann entropy is a valid analog of the thermodynamic entropy) use the entropy calculated from $S$ in thermodynamic calculations.}
	 
Expressed in words, Equation \eqref{dualstate} indicates that \textit{{entropy-type terms are related to correlations and vice versa}}. This kind of relation has already been exploited in a preliminary fashion to derive the Boltzmann distribution in a previous work by one of the authors \cite{Schmitz2020}. Moreover, it is also made plausible by the applications of entropy to entanglement \cite{HahnWK2020}, which is also a correlation.
All in all, the fraction of the boundary volume $\Phi_B $ has been identified as a further DoF and the extended internal energy now reads
\begin{equation}
		U= U(S,\Phi_B, V,\kappa,\vec r, \dotsc) \enspace \text{(internal energy of a quantized empty box)}.
		\label{newinternalenergyentropy}
\end{equation}
An important factor in this description is the finite thickness of the boundaries of the box leading to their finite volume. The caption of Equation \eqref{newinternalenergyentropy} already indicates a possible model of a quantized space. This model is further detailed in the following section.
	
\section{\label{chopping}Chopping the Box---Quantization of Space}
	This section introduces a model of a quantized space. Here, we do not claim that real space is de facto quantized but only present a model assumption. One of the major benefits of this perspective is the possibility of generating ratios of volume/area, area/line, volume/point, etc., which are all dimensionless numbers under the assumption of a quantized space, which we take here to imply that areas, lines, and points are three-dimensional objects that are very thin in one or more directions. Any volume will be an integral multiple of an elementary volume in this case. The model is based on the assumption that any geometric object is a three-dimensional (3D) object. {There is} extensive work in both philosophy and mathematics that questions, in one way or another, the geometric concepts of point, line, and surface figuring in traditional formulations of Euclidean synthetic geometry. Whitehead instigated a program in which points are defined by abstraction from “3-D” regions (see \cite{GerlaMirandam} for an overview). Roeper \cite{Roeper} regards points as locations in space but not as parts or elements of space or the primary bearers of spatial properties and relations. Johnstone \cite{Johnstone} reviews the development of pointless geometry in mathematics.

 Surfaces, lines (edges), and points (vertices) all are geometric objects that have a finite 3D extension on this view. Surfaces are small in one dimension (with a thickness of $\eta$), lines are small in two directions (with a cross-section of $\eta^2$), and vertices/points are small in all three directions (with a finite volume of $\eta^3$). The smallest thickness is denoted by $\eta$. From this perspective, any volume---and any geometric object---can be composed of non-overlapping elementary volume elements, each having the volume $\eta^3$. By analogy with regular numerical grids, these elementary volume elements are called {$voxels$.} 
 The overall box can then be chopped into a number $N_{\mathrm{box}}$ of voxels. These voxels could be floating around in some space individually. However, as they are all parts of a single geometric object---the box---each voxel is connected to at least one other voxel. Requiring the geometric object to be a box puts further constraints on their arrangement, allowing them to be classified into specific types of geometric sub-objects such as the voxels forming faces, those forming edges, the vertex voxels, and the voxels forming the bulk interior.
	
\begin{figure}[H]
		\includegraphics[width=0.48\textwidth]{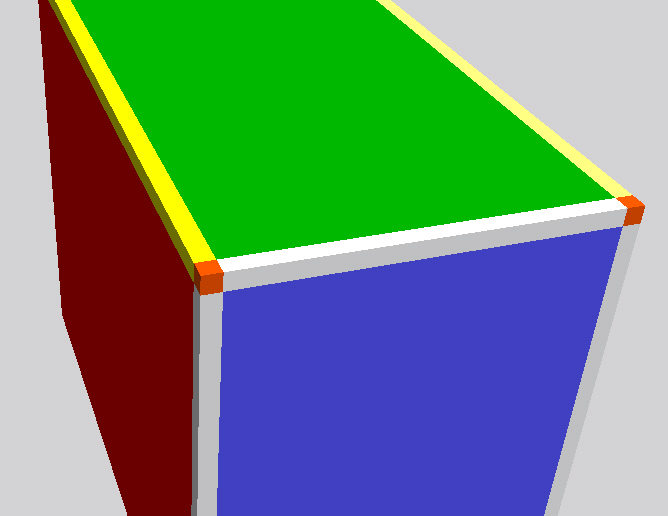}
		\caption{Visualization of the different voxel types: face voxels of the different faces (blue, green, red), edge voxels (light gray, yellow), and vertex voxels (orange).}
		\label{fig:Voxels}
	\end{figure}

	The total volume of a box is composed of its voxels, $N_{\mathrm{voxels}}$, each having the volume of an elementary volume element $\eta^3$.
	\begin{equation}
		V^{\text{total}}=N_{\text{voxels}}\eta^3. 
	\end{equation}
	There is no need to consider each of the voxels {themselves} as a DoF. They can be \linebreak  classified~into: 
	\begin{itemize}
		\item bulk voxels,
		\item face voxels,
		\item edge voxels,
		\item vertex voxels.
	\end{itemize}
	They then behave coherently (Figure \ref{fig:Voxels}). 
	
	In the first step, they are classified into voxels of an interior volume $V_I$ and voxels of a boundary volume $V_B$. The total volume is then found to be 
	\begin{equation}
		V^{\text{total}}=V^{\text{interior}}+V^{\text{boundary}}. 
		\label{totalvolume}
	\end{equation}
	{Again, we assume (as is thermodynamically reasonable \cite{Schmitz2003}) that the boundaries have a finite extension in three dimensions and therefore a non-zero volume.} The boundary volume itself is composed of different parts, faces, edges, and vertices, which are also volumes. It is given by
	\begin{equation}
V^{\text{boundary}}=V^{\text{faces}}+V^{\text{edges}}+V^{\text{vertices}}
	\end{equation}
	where $V^{\text{faces}}$ is the volume of the faces, $V^{\text{edges}}$ is the volume of the edges, and $V^{\text{vertices}}$ is the volume of the vertices. The total volume is thus given by\begin{equation}
		V^{\text{total}}=V^{\text{interior}}+V^{\text{faces}}+V^{\text{edges}}+V^{\text{vertices}}.  
	\end{equation}
		To account for the small extension in one, two, or all three dimensions of the faces, edges, and vertices, respectively, a small but finite length $\eta$ is introduced. This allows the recovery of the classical description of area $A$, length $L$, and points $P$ {in the form}
	
	\begin{equation}
		V^{\text{total}}=V^{\text{interior}}+\eta A^{\text{faces}}+\eta^2 L^{\text{edges}}+\eta^3 P^{\text{vertices}}
		\label{vtotal}
	\end{equation}
	where $A^{\text{faces}}$ is the area of the faces, $L^{\text{edges}}$ is the length of all edges, and $P^{\text{vertices}}$ is the number of vertices. Dropping the superscripts and denoting the total volume by $V_T$, Equation \eqref{vtotal} can be written as
	\begin{equation}
		V_T=V+\eta A+\eta^2L +\eta^3 P.
		\label{vtotal2}
	\end{equation}
From this, we can define the volume of the boundary as 
	\begin{equation}
		V^{\text{boundary}}=\eta(2ab+2bc+2ac)+ \eta^2(4a+4b+4c) + 8\eta^3.
	\end{equation}
	The individual entities are all volume contributions in this case. The numbers of faces, edges, and vertices are easily identified as 6, 12, and 8, respectively. The total volume of the box is obtained by adding the interior volume $V=abc$ to the above boundary volume~$V^{\text{boundary}}$:
	\begin{equation}
		V^{\text{box}}=abc + \eta(2ab+2bc+2ac)+ \eta^2(4a+4b+4c) + 8\eta^3.
	\end{equation}
	Further, it can be argued that only one-half of each of the surface areas, one-quarter of each line/edge, and one-eighth of each vertex belong to the volume of the box itself, whereas the other fractions are part of adjacent boxes. The situation is similar to the definition of a unit cell in crystallography \cite{UnitCell} {(although here, we physically consider an isolated box rather than a lattice, as is common in solid-state physics, the discussion of adjacent boxes primarily serves to illustrate the geometry)}. Any face is shared with another unit cell, each edge with four other unit cells, and each vertex with eight neighboring unit cells. These considerations then yield:
	\begin{equation}
		V^{\text{box}}=abc + \eta(ab+bc+ac)+ \eta^2(a+b+c) + \eta^3.
		\label{vbox}
	\end{equation}
	Equation \eqref{vbox} clearly recovers the classical---sharp interface---description for $\eta \rightarrow 0 $:
	\begin{equation}
		\lim_{\eta \to 0} V^{\text{box}}=abc .
	\end{equation}
		Equation \eqref{vtotal2} can further be normalized by dividing by the total volume $V_T$:
	\begin{equation}
		1 =\frac{V}{V_T}+\frac{\eta A}{V_T}+\frac{\eta^2L}{V_T} +\frac{\eta^3P}{V_T}.
	\end{equation}
	Summarizing this section on quantization, the most important assumption is that any geometric object, whether volume, line, area, or point, has a finite volume. The sharp interface counterparts of areas with size $A$, lines with length $L$, and a number of points $P$ (which do not have a volume but also have a different physical dimension) can be related to the quantized description as follows:
	\begin{eqnarray}
		V_{\text{Area}} = \eta A\nonumber,\\
		V_{\text{Line}} = \eta^2 L,\\
		V_{\text{Point}} = \eta^3 P\nonumber.
	\end{eqnarray}
	A total energy can be assigned to this system of volume fractions by multiplying \linebreak  Equation~\eqref{vtotal2} with a pressure $p$. This generates notions of interfacial energy, line energy, and even point energy:
	\begin{equation}
		pV_T=pV + p\eta A + p\eta^2 L +p\eta^3P.
		\label{multipliedbypressure}
	\end{equation}
	The classical notions of interfacial energy $\sigma$ and line tension $\alpha$ with dimension energy per area and energy per length, respectively, and the point energy $\chi$ can be obtained by the following assignments
	\begin{equation}
		\sigma := p\eta , \enspace \alpha := p\eta^2 , \enspace \text{and} \enspace \chi:=p\eta^3.
		\label{multipliedbypressure2}
	\end{equation}\vspace{-12pt}	\begin{equation}
		pV_T=pV + \sigma A + \alpha L +\chi P.
		\label{ClassicalInterfacialEnergy}
	\end{equation}

\section{Applications of the Quantized Box Model}

The following subsections illustrate some implications of the model framework for a quantized box, as outlined in the previous Section \ref{chopping}.  
	\subsection{\label{ofgeometric}Thermodynamics of Geometric Objects}
	
	The volume of a simple \textit{{sphere}} is described by a single parameter, i.e., a single degree of freedom: the value of its radius $r$. Any change in radius directly results  in a change in the volume $\vert V \vert$ of this fully symmetric object. The condition of 3D rotational symmetry thus fixes all other DoFs and only leaves the radius unconstrained:
	\begin{equation}
		V=V(r).
	\end{equation}
	A sphere is characterized by its interior volume and its surface/boundary. The boundary ``area'' $A$ has already been described here as a finite volume that is thin in one dimension with a thickness $\eta$. This approach has been further explained in Section \ref{chopping}. The total volume $V^{\text{total}}$ of the sphere is given by the sum of its interior volume $V^{\text{interior}}$ and its boundary volume $V^{\text{boundary}}$, i.e.,
	\begin{equation}
		V^{\text{total}}=V^{\text{interior}}+V^{\text{boundary}}=V+\eta A.
	\end{equation}
	The thickness of the boundary is considered to be finite with a value $\eta$. Note that the size of a sphere cannot be altered without altering its surface area and surface volume. 
	The internal energy $U$ then can be written as
	\begin{equation}
		U(V(r))= TS-pV+\sigma A
	\end{equation}
	where $\sigma$ is the interfacial energy. The parameter $\sigma$ has the dimension of energy/area compared to pressure with the dimension of energy/volume. For isothermal and reversible ($\dif S=0$) processes, this leads to the differential
	\begin{equation}
		\dif U(V(r))=-p\dif V+\sigma\dif A .
	\end{equation}
	At equilibrium, the variation in internal energy $U$ will be 0:
	\begin{equation}
		\dif U=0
	\end{equation}
	This then yields
	\begin{equation}
		0=-p\dif V+\sigma\dif A \rightarrow \enspace\enspace p\dif V=\sigma \dif A \enspace\enspace \rightarrow \enspace\enspace p=\sigma\frac{\dif A}{\dif V}.
	\end{equation}
	Inserting the differentials $\dif A=8\pi r\dif r$ and $\dif V= 4\pi r^2\dif r$ finally yields
	\begin{equation}
		p=\frac {2\sigma}{r}.
	\end{equation}
	This is the well-known Young--Laplace equation relating the pressure of a gas in a soap bubble to its radius \cite{YoungLaplace}. The radius of a sphere is not a degree of freedom that can be varied while  the volume is kept constant. A cylinder is an example of a geometric object that allows for a variation of one of its DoFs while keeping its total volume constant.
	
	{\label{cyli}}The volume of a \textit{{cylinder}} with a rotational symmetry along one axis exhibits at least two DoFs: its length $l$ and its radius $r$. A change in one or more of these DoFs may---or may not---change the value of the overall volume of the cylinder. In case the cylinder is elongated in the axial dimension, it may preserve the volume by a contraction in the radial dimension. This is related to the Poisson effect in mechanics, which is measured by Poisson's ratio $\nu$ relating the transverse and longitudinal strain \cite{Poisson2}. The differential of the volume $V(l,r)$ reads
	\begin{equation}
		\dif V = \pdif{V}{l}\dif l + \pdif{V}{r}\dif r .
		\label{cylinderdifferential}
	\end{equation}
	In the case of a conserved volume, i.e., $\dif V=0$, Equation \eqref{cylinderdifferential} results in the condition
	\begin{equation}
		\pdif{V}{l}\dif l =- \pdif{V}{r}\dif r .
	\end{equation}
 For a cylinder ($V = \pi r^2 l $), we obtain 
 \begin{equation}
  \pdif{V}{l}\dif l = \pi r^2 \dif l  \enspace\enspace \textit{and} \enspace \enspace  -\pdif{V}{r}\dif r = -2\pi r l \dif r
 \end{equation}
 Division by $\pi$ and $r$ leads to
 \begin{equation}
  r \dif l  = -2 l \dif r
 \end{equation}
 and eventually to
 \begin{equation}
  \frac{\dif l}{l}  = -2 \frac{\dif r}{r} \rightarrow \enspace -0.5 \frac{\dif l}{l}  = \frac{\dif r}{r} \rightarrow  -\nu\frac{\dif l}{l}  = \frac{\dif r}{r}
\end{equation}
 This reflects the Poisson effect for $\nu = 0.5$, in the case of isotropic, volume-conserving materials and small elongations.
	
	{\label{toab}}
	
	Several aspects of a \textit{{box}} make a deeper discussion interesting. In contrast to spheres and cylinders, boxes can be used to densely tessellate a 3D universe. Furthermore, they exhibit a full 3D anisotropy (in contrast to a cube). Compared with a sphere and a cylinder, a box exhibits different kinds of characteristic \textit{{boundary elements}}, namely faces, lines, and vertices.	The volume $V$ of a box is anisotropic in three dimensions with length $a$, width $b$, and height $c$.

	\subsection{Dimensionless Entities}
	
	One of the interesting aspects of expressing all types of geometric objects as volumes, as described in Section \ref{chopping}, is the fact that dimensionless parameters can be used. This especially holds for ratios of only two entities, where everyone knows the famous dimensionless relation between the perimeter and diameter of a circle, given by the \textit{{constant}}
	\begin{equation}
		\pi=\dfrac{\text{perimeter}}{\text{diameter}}.
	\end{equation}
	An important example of a dimensionless \textit{{constant}} with multiple entities is the fine-structure constant $\alpha$ \cite{CODATA2018}, which, as has been argued, is possibly related to discretized circles and polygons \cite{alpha137}. It is given by 
	\begin{equation}
		\alpha=\dfrac{1}{137.035999084(21)} = \dfrac{e^2}{4\pi\epsilon_0\hbar c}
	\end{equation}
	where $e$ is the electron charge, $\epsilon_0$ is the vacuum permittivity, $\hbar$ is the reduced Planck constant, and $c$ is the speed of light. A variety of such dimensionless numbers, which may be \textit{{constants}} or \textit{{variables}}, can be defined. Some of them are discussed below.

	The ratio of ``area volume'' and ``point volume'' is { such } a dimensionless number. It~reads
	\begin{equation}
		S=\dfrac{\eta A}{\eta^3}=\dfrac{A}{\eta^2}.
	\end{equation}
	This is already close to the entropy formula for a black hole in the geometric version in Equation \eqref{geometric}. The factor $1/4$ is still missing. The required step
	\begin{equation}
		S=\dfrac{A}{\eta^2} \enspace\enspace \rightarrow \enspace\enspace S=\dfrac{A}{4 l_p^2}
	\end{equation}
	can be performed by assuming an entropy-type distribution of gradients (or contrast) in the transition region of thickness $\eta$, with $1/l_p$ denoting the maximum possible contrast gradient. A full derivation can be found in Refs.~\cite{Schmitz2018,Schmitz2020}. This expression then matches the Bekenstein--Hawking Equation in \eqref{geometric} for the entropy of a black hole \cite{Bekenstein1972}.
 
	\subsection{\label{squeezing}Squeezing the Box}
\begin{figure}[H]
		\includegraphics[width=0.48\textwidth]{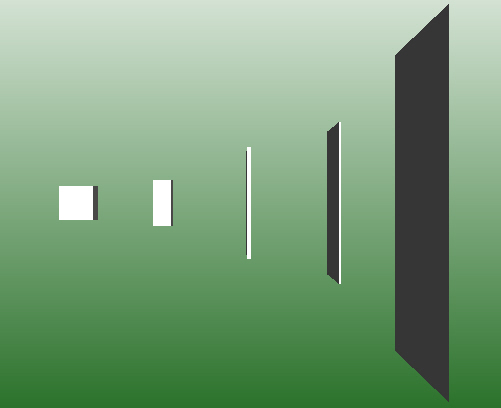}
		\caption{Visualization of a transition between a cube ($\cos(\theta) =1$) and a plane ($\cos(\theta) =0)$. The volume is kept the same in all cases.}
		\label{fig:Squeeze1}
	\end{figure}

	Two simple vectors that are very relevant for the description of an anisotropic box and the degree of its anisotropy are the vector of the space diagonal $\vec n$ and the wave vector $\vec k$. The wave vector is the \ZT{reciprocal vector}: %
	\begin{equation}
		\vec k = \left(\begin{array}{r} a^{-1} \\ b^{-1}  \\ c^{-1}\end{array}\right) \enspace\enspace \text{and} \enspace \enspace \vec n =\left(\begin{array}{r} a \\ b  \\ c\end{array}\right).
	\end{equation}
	A scalar can be defined as the scalar product of these vectors. Clearly,  this scalar product yields a constant denoting the dimension of the underlying vector space (3 in the present~case):

	\begin{equation}
		\vec{k} \cdot \vec{n} =3 \enspace \text{(number of spatial dimensions)}.
		\label{scalarproduct}
	\end{equation}
	The scalar product in Equation \eqref{scalarproduct} can be related to a multiparameter homogeneous function (see Section \ref{eulerh} on Euler homogeneity) describing the box. Simply multiplying both sides of Equation \eqref{scalarproduct} with the volume of the box gives, for the usual value $V(a,b,c)=abc$, the~result
	\begin{equation}
		abc\vec{k}  \cdot \vec{n} = 3abc. 
	\end{equation} 
	Writing the scalar product in components
	\begin{equation}
		abc\vec k = \left(\begin{array}{r} bc \\ ac  \\ ab\end{array}\right) \enspace\enspace \text{and} \enspace \enspace \vec n =\left(\begin{array}{r} a \\ b  \\ c\end{array}\right)
	\end{equation}
	would specify $V=abc$ be a homogeneous function in the case where the multiplication of the volume with $\vec{k}$ represents the derivative, which is obviously the case:
	\begin{equation}
		abc\vec k = \left(\begin{array}{r} \pdif{abc}{a}  \\ \pdif{abc}{b} \\ \pdif{abc}{c}\end{array}\right) 
		=: \pdif{V(x_i)}{x_i}.
	\end{equation}
	The multiplication of the scalar volume with the wave vector $\vec k$ obviously corresponds to a differentiation operation.
	Rewriting the scalar product according to well-known rules provides the specification of an angle $\theta$:
	\begin{equation}
		3=\vec{k}\cdot \vec{n} =\vert \vec k \vert \vert\vec n\vert \cos\theta.
		\label{angle}
	\end{equation}
	It can be directly inferred that the two vectors are parallel in the case of $\cos(\theta)=1  \textit{ and }\theta=0$ and perpendicular when $\cos(\theta) = 0$. The latter case of perfect perpendicularity clearly can never be reached by these two vectors.
	\begin{equation}
		\cos\theta=1 \leftrightarrow  \vec n \parallel \vec k, 
	\end{equation}
	\begin{equation}
		\cos\theta=0 \leftrightarrow  \vec n \perp \vec k. 
	\end{equation}
	The two vectors are parallel in the case where both vectors are linearly dependent, i.e., one of them is a scalar multiple of the other one:
	\begin{equation}
		\vec n \parallel \vec k \leftrightarrow  \vec n=scalar \cdot \vec k. 
		\label{scalarmultiple}
	\end{equation}
	It can be seen that this can be realized for \textit{cubes} where $a=b=c$ and using $a^2$ as the scalar in Equation \eqref{scalarmultiple}
	\begin{equation}
		a^2\left(\begin{array}{r} a^{-1} \\ a^{-1}  \\ a^{-1}\end{array}\right) =\left(\begin{array}{r} a \\ a  \\ a\end{array}\right).
		\label{lineardependent}
	\end{equation}
	As a special case, both vectors might be identical, i.e., $\vec k=\vec n$. This then yields a \textit{unit cube} with $a=1$, for which Equation \eqref{lineardependent} reads
	\begin{equation}
		1^2\left(\begin{array}{r} 1 \\ 1  \\ 1\end{array}\right) =\left(\begin{array}{r} 1 \\ 1  \\ 1\end{array}\right).
	\end{equation}
	A four-dimensional (4D) box in this context would generate a factor of $n_{\mathrm{dim}} = 4$. In the case where both 4-vectors are identical, i.e., $\vec k_4=\vec n_4$,  a 4D unit cube with $a=1$ is obtained:
	\begin{equation}
		\vec k_4=\left(\begin{array}{r} 1 \\ 1  \\ 1 \\ 1 \end{array} \right) =\left(\begin{array}{r} 1 \\ 1  \\ 1 \\1 \end{array}\right)=\vec n_4
	\end{equation}
	and 
	\begin{equation}
		\vec k_4 \cdot \vec n_4 = 4.
	\end{equation}
	Remember that to reach this description, all lengths in the 3D case have to be equal (i.e., $a=b=c$). In the case of extending to 4D, this would also hold for the fourth length $\tau$ (i.e., $\tau=a=b=c$). A limiting case---which cannot be reached but only asymptotically approached---is the case where both vectors are mutually perpendicular. This case ($\theta \approx \frac{\pi}{2}$) leads to an extremely compressed box, which is almost a plane. All cases with $\theta = (2n+1)\frac{\pi}{2}$ then correspond to planes and no longer represent a box. These mathematically sharp 2D planes obviously represent a further state of the system that complements the box with its finite volume. The perpendicularity can be expressed using the cross-product
	\begin{equation}
		\vert \vec k \times \vec n \vert= \vert \vec k\vert \vert\vec n\vert \sin(\theta).
	\end{equation}
	The angle $\theta$ is thus a parameter describing a transition from a ``cube'', i.e., an ideal 3D object, to a plane, i.e., an ideal 2D object, as visualized in Figure \ref{fig:Squeeze1}.
	
	\subsubsection{Special Relativity}
	
	A comparison can be made between two different boxes with the same volumes, but different shapes specified by the angle $\theta$. In particular, we compare an unspecified box with a special kind of box, namely a cube, and also a cube with a special cube, the unit~cube. 
 
 The two states are defined by a property $X$ and their difference can then be described as a positive factor $\gamma \geq 1$ similar to the notion of the Lorentz factor:
 
	\begin{equation}
		X_{\mathrm{box}}=\gamma X_{\mathrm{cube}} \enspace \enspace \rightarrow \enspace \gamma= \frac{X_{\mathrm{box}}}{X_{\mathrm{cube}}}
		\label{Lorentz1}.
	\end{equation}
	The scalar product definition holds for both the box and the cube
	\begin{equation}
		\cos(\theta)=\frac {\vec k \cdot \vec n} {\vert \vec k \vert\vert \vec n\vert}   ,
	\end{equation}
	\begin{equation}
		\vert \vec k \vert\vert \vec n\vert =\frac {\vec k \cdot \vec n} {\cos(\theta)} =\frac {n_{\mathrm{dim}}} {\cos(\theta)}.
		\label{tobesquared}
	\end{equation}
Equation \eqref{tobesquared} can be squared, which is useful for subsequent reasoning. Squaring Equation \eqref{tobesquared} gives
	\begin{equation}
		\vec k^2  \vec n^2=\frac {{n_{\mathrm{dim}}}^2} {\cos^2(\theta)}. 
	\end{equation}
We now use a prime symbol to indicate the parameters relating to the box, whereas the parameters of the cubes will not be primed. The square of the ratio of the properties, as defined in Equation \eqref{Lorentz1}, then reads ($X^2_{\mathrm{cube}} = \frac {n_{\mathrm{dim}}^2} {\cos^2(\theta)}$ and $X^2_{\mathrm{box}}= \frac {n_{\mathrm{dim}}^2} {\cos^2(\theta')}$):
	\begin{equation}
		\gamma^2=\frac {\cos^2(\theta)}{\cos^2(\theta')}.
		\label{gammasquared}
	\end{equation}
	Using the well-known relation 
	\begin{equation}
		1= \sin^2(\theta) + \cos^2(\theta),
		\label{standardidentity}
	\end{equation}
Equation \eqref{gammasquared} can be rewritten as
	\begin{equation}
		\gamma^2=\frac {\cos^2(\theta)}{1-\sin^2(\theta')}.
		\label{gammasquared2}
	\end{equation}
	Inserting the value 1 for $\cos^2(\theta)$---which is asserted for the case of a cube---into Equation~\eqref{gammasquared2} gives:
	\begin{equation}
		\gamma^2=\frac {1}{1-\sin^2(\theta')}
		\label{gammasquared3}
	\end{equation}
	which is the same as 
	\begin{equation}
		\gamma^2=\frac {c^2}{c^2-c^2\sin^2(\theta')}.
		\label{gammasquared4}
	\end{equation}
To obtain Equation \eqref{gammasquared4} from Equation \eqref{gammasquared3}, we have multiplied the numerator and denominator by a (in principle, arbitrary) quantity $c^2$. Giving $c^2$ a physical interpretation---e.g., the square of a velocity---with a physical dimension,  $c^2\sin^2\theta'$ must have the same physical dimension, enabling us to define a velocity $v$ and velocity ratio $\beta$ as
	\begin{equation}
		v^2:=c^2\sin^2\theta' \rightarrow \beta^2:=\sin^2(\theta')= \frac{v^2}{c^2}.
	\end{equation}
	Finally, we obtain the Lorentz factor in the familiar formulation
	\begin{equation}
		\gamma^2=\frac {1}{1-\beta^2}  \enspace\enspace \text{(Lorentz factor)}.
		\label{LorentzFactor}
	\end{equation}
	The variables have been deliberately named $\beta$ and $\gamma$ to emphasize the mathematical analogy with special relativity. 
	
Following the above ratio between the property of a box and the property of a cube, which has led to the specification of the Lorentz factor, the ratio between a cube and a unit cube is investigated. We now use primed values for the cube and non-primed values for the unit cube. Since both are cubes, both of them have $\cos(\theta) = 1$:
\begin{equation}
		1=\cos(\theta)=\frac {\vec k \cdot \vec n} {\vert \vec k \vert\vert \vec n\vert} = \frac {\vec k' \cdot \vec n'} {\vert \vec k' \vert\vert \vec n'\vert}.
\end{equation}

For cubes, all vector components of the two vectors are identical. We name them $\frac{l_p}{\lambda}$ for $\vec k$ and $\frac{\lambda}{l_p}$ for $\vec n$, where $l_p$ is a constant and $\lambda $ is a variable. Then, there exists a scalar variable $m^2(\lambda)$ scaling the two vectors
\begin{equation}
		\vec k = m^2(\lambda)\vec n
\end{equation}
For the unit cube, we have $\vec k$ = $\vec n$, which can only be realized for $\lambda= l_p$ and $m^2(l_p)=1$.
	\subsection{Evolution of the Box}
	The fact that the identity
	\begin{equation}
		n_{\text{dim}}=\vec{k}\cdot \vec{n} =\vert \vec k \vert \vert\vec n\vert \cos\theta 
	\end{equation}
	holds regardless of the choice of $a$, $b$, and $c$ makes it a sound and interesting starting point for further mathematical operations. If our box evolves with time $t$, the vectors $\vec k$ and $\vec n$ become time-dependent and we find that
	\begin{equation}
		n_{\text{dim}}=\vec k(t)\cdot \vec n(t) \enspace \enspace \text{for all }t.
		\label{scalarproduct_tdep}
	\end{equation}
	Similarly, the box parameters could depend on the position $\vec r$, implying that
	\begin{equation}
		n_{\text{dim}}=\vec k(\vec r) \cdot \vec n(\vec r)  \enspace \enspace \text{for all  } \vec r.
	\end{equation}
	The time derivatives of the space diagonal $\vec n$ and wave vector $\vec k$ can be calculated via their vector components as follows:
	\begin{equation}
		\dot {\vec k}=\frac{\dif}{\dif t}\vec k = -\left(\begin{array}{r}\dot a/ a^{2} \\ \dot b / b^{2}  \\ \dot c / c^{2}\end{array}\right) \enspace\enspace \text{and} \enspace \enspace \dot{\vec n}=\frac{\dif}{\dif t}\vec n =\left(\begin{array}{r} \dot a \\ \dot b  \\ \dot c\end{array}\right).
	\end{equation}
	Further, we can calculate the second derivatives with respect to time and find that
	\begin{equation}
		\ddot {\vec k}=\frac{\dif}{\dif t}\dot{\vec k} = -\left(\begin{array}{r}\ddot a/ a^{2} \\ \ddot b / b^{2}  \\ \ddot c / c^{2}\end{array}\right)+2\left(\begin{array}{r}{\dot a}^2/ a^{3} \\ {\dot b}^2 / b^{3}  \\ {\dot c}^2 / c^{3}\end{array}\right) \enspace\enspace \text{and} \enspace \enspace \ddot{\vec n}=\frac{\dif}{\dif t}\dot{\vec n} =\left(\begin{array}{r} \ddot a \\ \ddot b  \\ \ddot c\end{array}\right).
	\end{equation}
In the case where both vectors are solely functions of time, their full derivative, which is equal to the time derivative of $n_{\text{dim}}$, is equal to their partial derivative and reads
	\begin{equation}
		\frac {\dif n_{\text{dim}}}{\dif t}= \frac {\dif(\vec k \cdot \vec n)}{\dif t} \rightarrow 0 =  \vec n\cdot\dot{\vec k} + \vec k\cdot\dot{\vec n}.
		\label{fullderivative}
	\end{equation}
	The sum of these two scalar products clearly vanishes, indicating that one is the negative of the other. This allows for the identification of an anisotropic ``Hubble parameter'' $H_{\mathrm{aniso}}$ given by
	\begin{equation}
		H_{\mathrm{aniso}}:= -\vec n \cdot\dot{\vec k}= \frac{\dot a}{a}+\frac{\dot b}{b}+\frac{\dot c}{c}.
		\label{hubbleaniso}
	\end{equation}
	The name ``Hubble parameter'' has been chosen here because, for the special case of a cube $(a=b=c)$, this defines a parameter $H_{\mathrm{iso}}$, which up to a factor of 3, corresponds to the Hubble parameter $H$ entering into the Friedmann equations for the expansion of an isotropic universe \cite{Lemaitre1927}. In fact, Hubble parameters of the form in Equation \eqref{hubbleaniso} have already been used in anisotropic cosmological models \cite{AkarsuK2010}. For a cube, we recover the standard definition of the Hubble parameter given by
	\begin{equation}
		H= \frac{1}{3}{H_{\mathrm{iso}}} \enspace \text{with} \enspace {H_{\mathrm{iso}}}= 3\frac{\dot a}{a}.
	\end{equation}
	Starting with the time-dependent scalar product, Equation \eqref{scalarproduct_tdep}, the vectors can be evolved with a small time step $\dif t$ yielding
	\begin{equation}
		\vec k(t+\dif t)= \vec k(t) + \dot {\vec k} \dif t  \enspace \text{   and   } \enspace \vec n(t+\dif t)= \vec n(t) + \dot {\vec n} \dif t.
	\end{equation}
	Since their scalar product is invariant, we must have 
	\begin{equation}
		\vec n(t+\dif t) \cdot \vec k(t+\dif t)= (\vec k(t) + \dot {\vec k} \dif t)\cdot( \vec n(t) + \dot {\vec n} \dif t) = constant = n_{\text{dim}}.
	\end{equation}
	This results in a total of four terms
	\begin{equation}
		n_{\text{dim}}= \vec k\cdot\vec n + \vec n\cdot \dot {\vec k} \dif t + \vec k\cdot \dot{\vec n}\dif t +\mathcal{O}(\dif t^2).
		\label{4terms}
	\end{equation}
	This essentially corresponds to a Taylor expansion truncated to first order and thus reproduces the result of Equation \eqref{fullderivative}. More interesting is an expansion up to order $\dif t^2$, which by the same reasoning yields
	\begin{equation}
	 n_{\mathrm{dim}} = n_{\mathrm{dim}} + (\ddot{\vec{k}}\cdot\vec{n}+ 2\dot{\vec{k}}\cdot\dot{\vec{n}}+\vec{k}\cdot\ddot{\vec{n}})\dif t^2 + \mathcal{O}(\dif t^3). 
	\end{equation}
	The sum of all terms of second order, of course, also vanishes. For $a=b=c$ (the case of a cube), we would find the familiar result
	\begin{equation}
	 \dot{H}_{\mathrm{iso}}= \frac{\ddot{a}}{a} - \bigg(\frac{\dot{a}}{a}\bigg)^2 = \frac{\ddot{a}}{a} - H_{\mathrm{iso}}^2,   
	\end{equation}
	which is useful for cosmological calculations \cite{teVrugtHW2021}. Here, however, we have
	\begin{equation}
	\dot{H}_{\mathrm{aniso}}= \frac{\ddot{a}}{a} - \bigg(\frac{\dot{a}}{a}\bigg)^2 +\frac{\ddot{b}}{b} - \bigg(\frac{\dot{b}}{b}\bigg)^2+\frac{\ddot{c}}{c} - \bigg(\frac{\dot{c}}{c}\bigg)^2. 
	\end{equation}
On the other hand, we obtain
	\begin{equation}
		H^2_{\mathrm{aniso}} = \bigg(\frac{\dot a}{a}+\frac{\dot b}{b}+\frac{\dot c}{c}\bigg)^2 = - \dot {\vec k}\cdot \dot{\vec n} + 2\bigg(\frac{\dot a\dot b}{ab}+\frac{\dot b\dot c}{bc}+\frac{\dot c\dot a}{ca}\bigg).
	\end{equation}
Thus, we cannot calculate $\dot{H}_{\mathrm{aniso}}$ simply as a sum of terms of second order in derivatives and the square of $H_{\mathrm{aniso}}$ since squaring the anisotropic $H_{\mathrm{aniso}}$ of a box generates correlation terms between the expansion and contraction rates in the different directions. As a word of caution, one should understand that an anisotropic expansion rate $H$ in one direction would also allow for compensation by shrinkage in another direction (similar to the Poisson effect discussed in Section \ref{ofgeometric}).

The actual universe is observed to be anisotropic and the development of cosmological models incorporating such anisotropies is a topic of ongoing research \cite{sym12101741,teVrugtHW2021}. {Although the smallness of anisotropies in the cosmic microwave background (CMB) suggests a relatively homogeneous universe, at least at early times \cite{LarenaABKC2009}, the irrelevance of cosmological anisotropies for the expansion of the universe has not yet been demonstrated \cite{BuchertEtAl2015} and the fact that they should have at least some effect is a direct consequence of the mathematics of general relativity \cite{teVrugtHW2021}. In addition, the structure of the universe, and thus its inhomogeneity, has grown considerably since early times, which are captured by the CMB (see Refs.\ \cite{LarenaABKC2009,teVrugtW2019} for studies employing CMB data to analyze this problem).} It might also be interesting to exploit models allowing for expansion in one direction accompanied by simultaneous shrinkage in another direction similar to the Poisson effect.

	\subsection{\label{translating}Translating the Box}
	
	\subsubsection{\label{newton}Newton's Laws}
	
	 The thermodynamic description has yet to take into account the position of the box. Moving the box corresponds to changes in its centroid position $\vec{r}$. The thermodynamic potential, in this case, will, therefore, also be a function of  $\vec{r}$, indicating that this position is another DoF of the system. We still have an empty box with no particles inside ($N$=0) and the internal energy reads 
  \begin{equation}
		U(S,V,\vec r).
	\end{equation}
	It has the differential
	\begin{equation}
		\dif U = T\dif S -p\dif V +\pdif{U}{\vec{r}}\cdot\dif \vec{r}.
		\label{urdifferential}
	\end{equation}

Newton's second law can be deduced from thermodynamics by assuming that the position $\vec {r}$ of the centroid of the box is a function of time $t$. For a \textit{non-dissipative and isochoric process} (i.e., $\dif S=0$ and $\dif V=0$), the differential of $U$ reads
	\begin{equation}
		\dif U=\pdif{U}{\vec r}\cdot\pdif{\vec r}{t}\dif t  =:\dot E \dif t.
	\end{equation}
Here, the product of the two partial derivatives has been identified with the time derivative of an energy $E$ (that, for the moment, is simply $U$):
		\begin{equation}\label{eqn:Edot}
		-\dot E= -(\vec\nabla U) \cdot \dot{\vec r} := \vec F \cdot \vec v. 
	\end{equation}

Defining the momentum vector $\vec p$ via its differential as
 		\begin{equation}
			\dif \vec p := \vec F \dif t \enspace \rightarrow \enspace \vec F = \dot{\vec p}
			\label{DefMomentum}
		\end{equation}
yields
\begin{equation}
			-\dot E = \dot{\vec p}\cdot \vec v. 
			\label{dote3}
		\end{equation}
Finally, we can note that the energy should be conserved, i.e., $\dot E = 0$. Then, Equations \eqref{DefMomentum} and \eqref{dote3} give
		\begin{equation}
			\dot E = 0 \rightarrow \vec{F} = \vec 0 \rightarrow \dot{\vec p} = \vec{0} \rightarrow \vec p = constant \enspace \vee \enspace \vec{v}= \vec{0} \enspace \text{ (Newton's first law)}.   
			\label{newtonsfirstlaw}
		\end{equation}
Although these results are encouraging, two problems remain:
\begin{enumerate}
    \item So far, we cannot say anything about how the quantity $\vec p$, defined by Equation \eqref{DefMomentum}, is related to the velocity $\vec v$. Of course, Equation \eqref{newtonsfirstlaw} is only equivalent to the statement usually thought of as Newton's first law if $\vec p$ is proportional to the velocity.
    \item Energy conservation in a closed system should also hold for the more general case $\vec F \neq \vec 0$ (whereas Equation \eqref{eqn:Edot} suggests a direct dependence of $\dot{E}$ on $\vec{F}$, which motivated the first implication in Equation \eqref{newtonsfirstlaw}). A position-dependent potential does not lead to a violation of energy conservation.
\end{enumerate}
Both issues can be dealt with if we also allow the potential to depend on the velocity $\vec v$. Let us assume that the energy has a contribution $E_{\mathrm{kin}}$ (\ZT{kinetic energy}) that is simply added to the energy $E$ considered so far, which we denote with $E_{\mathrm{pot}}$ and refer to as the \ZT{potential energy}. The total energy $E$ is then given by
\begin{equation}
E = E_{\mathrm{pot}} + E_{\mathrm{kin}}.   
\label{ekinepot}
\end{equation}
Equation \eqref{dote3} implies that
\begin{equation}
\dot{E}_{\mathrm{pot}} = - \dot{\vec{p}}\cdot\vec{v}.
\label{dotepot}
\end{equation}
Imposing the condition $\dot{E} = 0$, Equations \eqref{ekinepot} and \eqref{dotepot} give
\begin{equation}
\dot{E}_{\mathrm{kin}} = \dot{\vec{p}}\cdot\vec{v}.  
\label{dotekin1}
\end{equation}
If we assume that $\dot{E}_{\mathrm{kin}}$ is some function of $\vec{v}$ (and nothing else), we find, in general, that
\begin{equation}
\dot{E}_{\mathrm{kin}} = \pdif{E_{\mathrm{kin}}}{\vec{v}}\cdot\vec a    
\label{dotekin2}
\end{equation}
where $\vec a = \dot{\vec{v}}$ is the acceleration. Comparing Equations \eqref{dotekin1} and \eqref{dotekin2}, we note that the simplest way to satisfy them simultaneously is to assume that $\dot{\vec{p}}$ is proportional to $\vec{a}$ and $\pdif{E_{\mathrm{kin}}}{\vec{v}}$ is proportional to $\vec{v}$. The proportionality constants can then simply be obtained from dimensional analysis. In both cases, the proportionality constant needs to have a dimension of mass such that we can choose the mass $m$ of the box. This yields the standard results
\begin{align}
E_{\mathrm{kin}} =& \frac{1}{2}m\vec{v}^2,\\
\vec{p} =& m \vec{v}\label{DefMomentum2}.
\end{align}
Combining Equations \eqref{DefMomentum} and \eqref{DefMomentum2} gives
\begin{equation}
\vec{F} = m \vec{a}\quad \text{ (Newton's second law)}.   
\end{equation}

In the absence of any force and for a constant mass, the momentum will be a constant and Newton's first law is obtained: 
	\begin{equation}
		\vec F =\vec{0} \rightarrow \vec p = constant \enspace \rightarrow  \enspace \vec v = constant \enspace\enspace \text{(Newton's first law)}.
	\end{equation}
    Note that we have simplified our approach here by obtaining Newton's first law simply as a special case of the second law. Although this is not uncommon in introductory presentations, it may not be seen as completely adequate from a philosophical \linebreak  perspective (see \cite{Brown2005}, p. 37). Strictly speaking, Newton's first law introduces inertial frames of reference in which Newton's second law then holds (in the absence of apparent forces).
    
	Newton's third law was already discussed in Section \ref{spgf}.
 
 Beyond the mere classical kinetic energy discussed above, other terms yielding a velocity term can also be part of the energy, for example, terms proportional to ${\vec r}^2$. This would be the lowest-order term in a Taylor expansion of the contribution $E_{\mathrm{pot}}$. We, therefore, find that
	\begin{equation}
		E= \frac{k}{2}{\vec r}^2 + \frac{m}{2}{\dot{\vec r}}^2.
	\end{equation}
	Executing the time derivative then yields
	\begin{equation}
		\dot E= k{\vec r}\cdot\dot{\vec r} + m{\dot{\vec r}}\cdot\ddot {\vec r}.
	\end{equation}
	Separating $\dot{\vec r}$ and renaming $\dot{\vec r}$ as $\vec v$ and $\ddot{\vec r}$ as $\vec a$ gives
	\begin{equation}
		\dot E= (k{\vec r} + m\ddot {\vec r})\cdot\dot{\vec r}=(k{\vec r} + m {\vec a})\cdot\vec v.
	\end{equation}

 A non-dissipative process (i.e., $\dot E = 0$) corresponds to $k{\vec r} + m {\vec a}=\vec{0}$ and/or $\vec v=\vec{0}$. For a non-vanishing velocity $\vec v \neq \vec{0}$, this then leads to the equation for a harmonic oscillator:
	\begin{equation}
		\vec{0}= k{\vec r} + m {\ddot{\vec r}}\enspace \enspace \text{(Harmonic Oscillator)}.
	\end{equation}	

\subsubsection{The Unruh Effect}	
	
Next, we demonstrate how translating the box can allow for a classical derivation of a relation formally resembling the definition of the Unruh temperature. To have a physical idea of how this can be achieved, imagine that a box with mass $m$ is at rest in a gravitational field with strength $g$ and has a certain potential energy. If we let it fall a distance $z$ and assume that \ZT{falling} is an adiabatic process, the energy $m g z$ is converted into the kinetic energy of the center-of-mass motion. Next, the box is stopped instantaneously, which implies that the kinetic energy is converted into heat. Overall, this constitutes a process in which the box has moved a distance $z$, during which time an amount $m g z$ of energy is converted into heat. The equivalence between assuming that the box is in a constant gravitational field and assuming that it is subject to a constant acceleration, which we call $\vec{a}$, can now be observed. Finally, we assume that this process takes place over a very small length $\dif \vec{r}$. Thus, the energy that is converted into heat is given by $m\vec{a}\cdot\dif\vec{r}$, and we find that
	\begin{equation}
		T\dif S=m\vec{a}\cdot\dif\vec{r}.
	\end{equation}
	This indicates a possible path to the Unruh temperature $T_{\text{Unruh}}$ \cite{Unruh}, which reads
	\begin{equation}
		k_BT_{\text{Unruh}}=\frac{\hbar}{2\pi c}\mid\vec{a}\mid. 
	\end{equation}
	We can continue to follow this path by replacing the mass $m$ of the particle with its Compton wavelength $\lambda$:
	\begin{equation}
		\lambda=\frac {\hbar}{2\pi m c} \enspace \leftrightarrow \enspace m =\frac {\hbar}{2\pi\lambda c}. 
	\end{equation}
	This replacement leads to
	\begin{equation}
		T\dif S=\frac {\hbar}{2\pi\lambda c}\vec{a}\cdot\dif\vec{r} \enspace\enspace  \text{(Differential Unruh Equation)}
	\end{equation}
	Under the reasonable assumption that the acceleration $\vec a $ is parallel to the displacement $\dif\vec r $, the scalar product turns into a simple product, 
	\begin{equation}
		T\dif S=\frac{\hbar}{2\pi c}a\frac{\dif r}{\lambda},
	\end{equation}
	where $a = \vert \vec a \vert$. This is almost identical to the formulation of the Unruh temperature and would be identical if the displacement were to match the Compton wavelength. Integrating the displacement over one Compton wavelength and assuming the acceleration $a$ to be constant over this distance yields
	\begin{equation}
		T\Delta S=\frac{\hbar}{2\pi c}a\\ \int_{0}^{\lambda}\frac{1}{\lambda}  \,\dif r =\frac{\hbar}{2\pi c}a.
	\end{equation}
	Identifying the resulting $\Delta S$ with the Boltzmann constant $k_B$,
	\begin{equation}
		k_B:=\Delta S,
		\label{definitionofkb}
	\end{equation}
	finally yields the Unruh equation in its familiar form
	\begin{equation}
		k_BT=\frac {\hbar}{2\pi c}a \enspace\enspace \text{(Unruh Equation)}.
	\end{equation}
	The Boltzmann constant $k_B$ in this case can be interpreted as a ``quantum'' of entropy defined by the amount of entropy generated by shifting a particle by exactly its Compton wavelength. {More precisely, the constant $k_B$ is defined by Equation \eqref{definitionofkb} and does not have to be numerically identical to the Boltzmann constant used in statistical mechanics (since, here, we are deriving an analogy {with} the Unruh equation and not the Unruh equation itself).}

\subsubsection{Position-Dependent Volume}
The deliberations in Section \ref{newton} are not affected by the fact that we consider a box rather than, say, a point mass. Let us now, as indicated in Section \ref{spgf}, assume that there is also a position-dependent volume, i.e., that $V$ changes if the box is moved. Assuming a non-dissipative process ($\dif S = 0$), this gives
\begin{equation}
\dif U =  -p\pdif{V}{\vec r} \cdot\dif \vec r+ \pdif{U}{\vec r}\cdot\dif \vec r+\pdif{U}{\vec{v}}\cdot\dif \vec{v},
\label{difuwithvolume}
\end{equation}
where we have used the definition of pressure (see Equation \eqref{pTmuDefinitions}). Exploiting the fact that $\dif U = 0$ in a closed system and using $\pdif{U}{\vec{v}} = \vec{p}$ and $\dot{\vec{v}}=\vec{a}$ (see Section \ref{newton}), dividing Equation \eqref{difuwithvolume} by $\dif t$ gives\begin{equation}
0 = \vec{v}\cdot \bigg(-p\pdif{V}{\vec r} + \pdif{U}{\vec r}  +m \vec{a}\bigg),    
\end{equation}
giving
\begin{equation}
m \vec{a} = \vec{F} = p\pdif{V}{\vec r} - \pdif{U}{\vec r}.   
\end{equation}
Hence, the position dependence of the volume creates an additional contribution to the total force.

This result can be given the following physical interpretation: the term $\pdif{U}{\vec r}$ corresponds to an explicit position dependence of $U$ arising from an external potential. For a point mass, a shift of the position, which leads to a change in the potential energy, must (due to energy conservation) be compensated for by an increase in the kinetic energy. Hence, the external force generates acceleration. Here, however, the energy coming from the external potential is transformed partly into center-of-mass kinetic energy and partly into internal energy, resulting from the change in volume. This is fully analogous to the well-known observation that a cylinder rolling down an inclined plane arrives at the bottom more slowly than a point of the same mass since for the cylinder, not all of the initial potential energy is converted into center-of-mass kinetic energy, as some of it is converted into rotational energy.

	\subsubsection{Uncertainty Relation}

Finally, we discuss the implications of the \ZT{discretized} view on space for the dynamic equations if we apply it to both space and time. Replacing $\dif t$ by $\Delta t$, Equation \eqref{dotekin1} reads
\begin{equation}
 \frac{\Delta E_{\mathrm{kin}}}{\Delta t}= \frac{\Delta \vec{p}}{\Delta t}\cdot\frac{\Delta \vec{r}}{\Delta t}.
 \label{dote3mod}
\end{equation}
Multiplying both sides of Equation \eqref{dote3mod} by $(\Delta t)^2$ gives an \ZT{uncertainty relation}, which is formally equivalent to an uncertainty relation that appears in quantum mechanics (although it is formulated for $E_{\mathrm{kin}}$ rather than $E$):
	\begin{equation}
		{\Delta E_{\mathrm{kin}}}{\Delta t} = {\Delta}\vec p\cdot{\Delta\vec r} \enspace\enspace  \text{(uncertainty relation)}.
		\label{uncertaintyrelation}
	\end{equation}
{Note that Equation \eqref{uncertaintyrelation} is not an uncertainty relation in the quantum-mechanical sense but simply a result of elementary mechanics that has the same formal structure as a certain quantum-mechanical result.}

	\section{\label{beyondtheempty}Beyond the Empty Box}
	A number of equations and relations have been discussed:
	\begin{itemize}
		\item all three laws of Newton (classical mechanics),
		\item the harmonic oscillator equation,
		\item the Unruh equation,
		\item an uncertainty relation.
	\end{itemize}

  In this article, a number of further observations and thoughts have emerged, which strike us as very interesting and will be discussed in future articles. These thoughts arose from relaxing the constraint of describing ``an empty box'' and are briefly mentioned below. They refer to \textit{{oriented surfaces}} (which takes us beyond a simple box), \textit{{twisting and shearing the box}} (dropping the rectangular characteristics of a box), and \textit{{filling the box}} (dropping the constraint of an ``empty'' box).
	\subsection{Oriented Surfaces}
\begin{figure}[H]
		\includegraphics[width=0.62\textwidth]{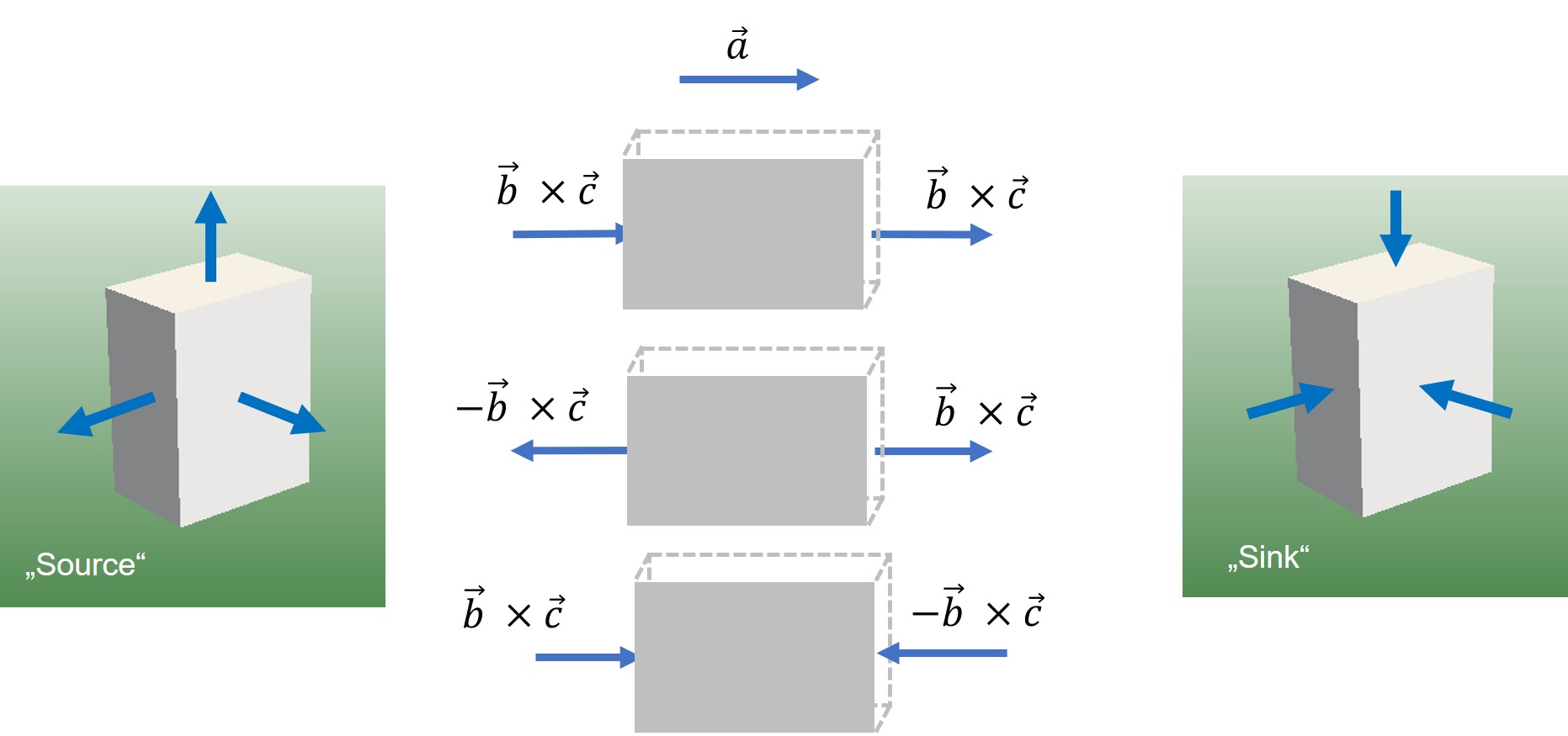}
		\caption{Source and sink as states of a box}
		\label{fig:SourceSink}
	\end{figure}

	The classical relation describing the charge of a capacitor
	\begin{equation}
		Q=CU_{\mathrm{electrostatic}}
	\end{equation}
	with the capacity $C$ and the voltage $U_{\mathrm{electrostatic}}$---the difference in the electrostatic potential---sustains a direct scalar relation between the charge $Q$ and the surface $A$ of a capacitor, given~by
	\begin{equation}
		Q=\epsilon_0\epsilon_r \frac{A}{d}U_{\mathrm{electrostatic}}
		\label{capacitor}
	\end{equation}
	where $\epsilon_r$ is the relative permittivity and $d$ is the distance between the plates. Equation \eqref{capacitor} can be re-written under certain assumptions as a scalar product of the oriented surface $\vec A$ and the gradient of a potential energy (the electrostatic potential) $\vec\nabla U$:
	\begin{equation}
		Q=\epsilon_0\epsilon_r \vec A\cdot\vec\nabla U_{\mathrm{electrostatic}}.
		\label{capacitor2}
	\end{equation}
	Equation \eqref{capacitor2} suggests a strong relation between the charge and both the area and area orientation. An approach to describing ``charges'' and electrostatics might thus arise from considering the \textit{{orientations of the faces}} of the box as an additional degree of freedom, where a ``positive charge'' would correspond to a box with all the normals of the faces pointing ``inwards'', whereas a negative charge would correspond to a box with all these normals pointing ``outwards'' (see Figure \ref{fig:SourceSink}).\vspace{-9pt}
	Similar interpretations of a quantum sink and quantum source are the philosophical foundations of the Elementary Multiperspective Material Ontology (EMMO) described in Ghedini et al. \cite{Ghedini2022} and can also be found in the cell method (Figures 12 and 13 in \cite{Ferretti2013}). 
	
	\subsection{Shearing and Twisting the Box}
	The present article did not exploit the spade product describing a volume element in its full depth but was limited to a box with parallel faces mutually perpendicular to each other. Considering the angles between non-perpendicular faces as further DoFs can be expected to result in the description of ``shear forces''. Twisting the two parallel faces of a box by some angle will relax the constraint of ``being a box'' and can be expected to lead to the description of a ``torque''.
	
	\subsection{Filling the Box}
	Filling the box will, generally, allow addressing (i) a box containing a single particle, and (ii) a box filled with massless particles. The model of a particle in a box is important in quantum mechanics as it is, in contrast to most other models, exactly solvable \cite{AtkinsQM} and it will be interesting to make comparisons. The derivation of the well-known Planck formula for the energy spectrum of black-body radiation also started with considering photons confined in a box with multiples of their wavelengths matching the size of the box \cite{planck1915}. The additional 2D DoF identified in Section \ref{squeezing} on \textit{{squeezing the box}} might represent an interesting starting point for the description of photons.

	
	\section{Summary and Outlook}
The present article started with the idea of identifying the ``hidden'' degrees of freedom (DoF) of an empty box. In classical thermodynamic potentials, the only degree of freedom would be the value of the volume $V$ of the box. Additional degrees of freedom were identified in this article, all leaving the volume value $V$ unchanged but relating to either a change in the shape or position of the box. Considering these DoFs as functional dependencies and performing the classical partial derivatives of the internal energy yielded a number of interesting relations. In particular, terms and equations corresponding to black-hole entropy \cite{Schmitz2018}, the Unruh temperature, and an uncertainty relation, as well as classical mechanics, could be identified. The thermodynamic approach pursued in this article demonstrates that familiar aspects of gravity, relativity, quantum mechanics and even quantum field theory can be obtained from a common thermodynamic perspective. The question arises of whether the same approach might  be further extended to also describe phenomena of electrostatics and electrodynamics. Thermodynamics in solid-state physics addressing cross-phenomena, e.g., between electrostatic fields and temperature, are expected to provide guidance here \cite{Liu2022}. Note that in the present article, special relativity can be considered a cross-phenomenon between the respective DoFs of ``squeezing the box'' and ``translating the box'', eventually leading to compression during translation resembling the relativistic length contraction. In our framework, forces have, in some cases, been identified as corresponding to correlations. Only binary correlations of type $\Phi_i\Phi_j$ have been exploited in the present article. It is anticipated that weak and strong interactions are related to higher-order ternary (i.e., $\Phi_i\Phi_j\Phi_k$) and quaternary (i.e., $\Phi_i\Phi_j\Phi_k\Phi_l$) correlations, raising a spectrum of options for a possible ``thermodynamics of everything'' (ToE).



	\vspace{6pt} 
	\authorcontributions{Conceptualization, G.J.S.; methodology, G.J.S.; software, not applicable; investigation, G.J.S. and L.E.; resources, G.J.S.; writing---original draft preparation, G.J.S., M.t.V., T.H.W., L.E., and P.N.; writing---review and editing, G.J.S., P.N., M.t.V., T.H.W., L.E., and R.W.; visualization/graphics, L.E.; supervision, G.J.S.; project administration, G.J.S.; funding acquisition, G.J.S. All authors have read and agreed to the published version of the manuscript.}
	
	\funding{M.t.V.\ thanks the Studienstiftung des deutschen Volkes for the financial support. G.J.S.\ and R.W.\ were funded by the Deutsche Forschungsgemeinschaft (DFG, German Research Foundation) -- Project-ID 39062112 -- EXC 2023; Project-ID 433682494 -- SFB 1459.}
	
	\institutionalreview{Not applicable}

	\dataavailability{No new data were created or analyzed in this study. Data sharing is not applicable to this article.} 
	
	\conflictsofinterest{The authors declare no conflicts of interest.}

	\appendixtitles{yes}
	\appendixstart
	\appendix
\section[\appendixname~\thesection. Legendre Transformations]{Legendre Transformations}\label{Legendre}
	The internal energy potential $U$ can be transformed into other potentials expressed in other---intensive or extensive---variables via the Legendre transformation
	\begin{equation}
		U_X := U - \pdif{U}{X} X, \quad X \in (S,V,N)
		\label{LegendreTransformUx}
	\end{equation}
	Examples are the Helmholtz energy $A$ or the enthalpy $H$:
	\begin{equation}
		A:= U_S, \enspace  H:= U_V . 
	\end{equation}
	\textls[-25]{From the outset, the derived potentials share the free variables of the parent potential $U(S, V, N)$, but there is another possibility with a simpler (so-called canonical) description, namely that the transformed variable is replaced by the corresponding partial derivative.~Hence,}
	\begin{equation}
		A = A\bigg(\pdif{U}{S},V,N\bigg), \enspace  H = H\bigg(S,\pdif{U}{V},N\bigg)
	\end{equation}
	or, after inserting the definitions in Equation \eqref{pTmuDefinitions},
	\begin{equation}
		A = A(T,V,N), \enspace  H = H(S,-p,N)
	\end{equation}
	These canonical forms are derived using two important  features of the partial differentiation of a Legendre transform with more than one variable:
	\begin{equation}
		\left(\pdif{U_X}{Y}\right)_{\pdif{U}{X}} = \left(\pdif{U}{Y}\right)_{X} ,
		\label{LegendreCommutativeDiff}
	\end{equation}
	\begin{equation}
		\left(\pdif{U_X}{\left(\pdif{U}{X}\right)_{Y}}\right)_{Y} = -X .
		\label{LegendreReflexiveDiff}
	\end{equation}
	Applied to $A$, after substituting for Equation \eqref{pTmuDefinitions}, we can write
	\begin{equation}
		\left(\pdif{A}{V}\right)_{T,N}  = \left(\pdif{U}{V}\right)_{S,N} := -p ,
	\end{equation}
	\begin{equation}
		\left(\pdif{A}{T}\right)_{V,N} = -S .
	\end{equation}
	The enthalpy $H$ and all the other imaginable Legendre transforms of $U$ follow the same pattern.
	
	The above derivative properties imply that the Legendre transformation, when expressed in canonical variables, is (i) commutative, and (ii) reversible, or, to continue our line of thought using $A$ as an example, we can calculate the continued transformation of $A_V =: G$ (called Gibbs energy) directly from $U$
	\begin{equation}
	    A_V := A - \pdif{A}{V} V = A - \pdif{U}{V} V := U - \pdif{U}{S} S - \pdif{U}{V} V =: U_{SV},
	\end{equation}
	and also illustrate how the repeated transformation of $S$ and $T$ in $U(S,V,N)$ and $A(T,V,N)$, respectively, returns the original potential $U(S,V,N)$ after three cycles:
    \begin{align}
	    U(+S,V,N)_{+S} & =: A(+T,V,N) \\
	    A(+T,V,N)_{+T} & =: U(-S,V,N) \\
	    U(-S,V,N)_{-S} & =: A(-T,V,N) \\
	    A(-T,V,N)_{-T} & =: U(+S,V,N) .
	\end{align}
	Positive and negative signs are deliberately used to illustrate the implications of the derivative property in Equation \eqref{LegendreReflexiveDiff}.
	
	The Legendre transformation defined in Equation \eqref{LegendreTransformUx} is linear and using a linear operator on a first-order Euler homogeneous function gives a unique result. The total transformation of $U$ gives the zero potential 
	\begin{equation}
		U_{XYZ} = 0; \quad (X\ne Y\ne Z) .
	\end{equation}
	Linearity also implies the existence of inversion and from Equation \eqref{InternalEnergy}, we can easily formulate several equivalent families of potentials, which can all undergo a Legendre transformation of the same kind we have seen above:
	\def\HWT{\makebox[\widthof{$-p$}][c]{$T$}}%
	\def\HWmu{\makebox[\widthof{$-p$}][c]{$\mu$}}%
	\begin{align}
	S(U,V,N) & = +\frac{1}{\HWT}U - \frac{-p}{\HWT}V - \frac{\HWmu}{\HWT}N , \\
	V(S,U,N) & = - \frac{\HWT}{ -p}S + \frac{1}{ -p}U - \frac{\HWmu}{ -p}N , \\
	N(S,V,U) & = - \frac{\HWT}{\HWmu}S - \frac{ -p}{\HWmu}V + \frac{1}{\HWmu}U .
	\end{align}
	The same inversion principle also applies to each of the Legendre transforms so, overall, there are many possibilities for expressing the state of a thermodynamic system.


\section[\appendixname~\thesection. Relation between Dual-State Entropy and Boltzmann Entropy]{Relation between Dual-State Entropy and Boltzmann Entropy}\label{boltzmann}

Here, we discuss in more detail why, from the perspective of statistical mechanics, it is reasonable to define the entropy via Equation \eqref{dualstate} if we specify the macrostate based on interior and boundary fractions.

An illustrative simple example is a system of three spins in one dimension that can be up ($\uparrow$) or down ($\downarrow$). All spins can flip randomly and all configurations have the same energy. We then describe this dual-state system in terms of the fraction of interior $\Phi_I$ (corresponding to the number of boundaries between spins pointing in the same direction) and boundary $\Phi_B$ (corresponding to the number of boundaries between spins pointing in opposite directions) parts, i.e., the macrostate is specified by $\Phi_I$ and $\Phi_B$. The system will evolve toward the most likely macrostate, i.e., the one revealing the largest number of possible microscopic realizations. This is the state with the largest Boltzmann entropy. The maximum of the dual-state entropy in Equation \eqref{dualstate} is reached when $\Phi_I = \Phi_B = 1/2$. If Equation \eqref{dualstate} is an appropriate expression for the entropy (assuming that the system evolves to the maximum entropy state), the macrostate $\Phi_I = \Phi_B = 1/2$ should also be the microscopically most likely one, i.e., the one with the largest Boltzmann entropy.
The possible configurations are 
\begin{center}
	$
	\uparrow\uparrow\uparrow,
	\enspace   \uparrow\uparrow\downarrow,  \enspace 
	\uparrow\downarrow\uparrow,  \enspace 
	\downarrow\uparrow\uparrow,
	\enspace  \uparrow\downarrow\downarrow,\enspace
	\downarrow\downarrow\uparrow,\enspace
	\downarrow\uparrow\downarrow,
	\enspace   \downarrow\downarrow\downarrow$.
\end{center}
As can be seen, there are two configurations with $\Phi_I=1$ and $\Phi_B=0$ ($\uparrow\uparrow\uparrow$ and $\downarrow\downarrow\downarrow$), two configurations with $\Phi_I=0$ and $\Phi_B=1$ ($\uparrow\downarrow\uparrow$ and $\downarrow\uparrow\downarrow$), and four configurations with $\Phi_I=\Phi_B=1/2$ ($\uparrow\uparrow\downarrow$, $\downarrow\uparrow\uparrow$, $\downarrow\downarrow\uparrow$, and $\uparrow\downarrow\downarrow$). Hence, the macrostate $\Phi_I=\Phi_B=1/2$ is indeed the one with the largest number of microscopic realizations and thus the maximum of the Boltzmann entropy.

\begin{adjustwidth}{-\extralength}{0cm}


\reftitle{References}


\PublishersNote{}
\end{adjustwidth}
\end{document}